\documentclass[prb,showpacs,twocolumn]{revtex4}%
\usepackage{graphicx}
\usepackage{dcolumn}
\usepackage{bm}
\usepackage{amsmath}
\usepackage{amsfonts}
\usepackage{amssymb}%
\usepackage{color}%
\providecommand{\U}[1]{\protect\rule{.1in}{.1in}}
\begin{document}

\title{
Derivation of Static Low-Energy Effective Models by \textit{ab initio} Downfolding Method without Double Counting of Coulomb Correlations: \\
Application to SrVO$_3$, FeSe and FeTe
}

\author{Motoaki Hirayama,$^{1,3}$ Takashi Miyake,$^{2,3}$ and Masatoshi Imada$^{1,3}$}

\affiliation{$^1$Department of Applied Physics, University of Tokyo, 7-3-1 Hongo, Bunkyo-ku, Tokyo 113-8656, Japan}
\affiliation{$^2$Nanosystem Research Institute, AIST, Tsukuba 305-8568, Japan}
\affiliation{$^3$CREST, JST, 7-3-1 Hongo, Bunkyo-ku, Tokyo 113-8656, Japan}
\date{October 11, 2012}

\begin{abstract}
Derivation of low-energy effective models by a partial trace summation of the electronic degrees of freedom far away from the Fermi level, called downfolding, is reexamined. 
We propose an improved formalism free from the double-counting of electron correlation in the low-energy degrees of freedom.
In this approach, the exchange-correlation energy in the local density approximation (LDA) is replaced with the constrained self-energy corrections defined by the sum of the contribution from eliminated high-energy degrees of freedom, $\Sigma^H$ and that from the frequency-dependent part of the partially screened interaction, $\Delta\Sigma^L$.  
We apply the formalism to SrVO$_3$ as well as to two iron-based superconductors, FeSe and FeTe.
The resultant bandwidths of the effective models are nearly the same as those of the previous downfolding formalism because of striking cancellations of $\Sigma^H$ and $\Delta\Sigma^L$.
In SrVO$_{3}$, the resultant bandwidth of the effective low-energy model is $2.56$ eV (in comparison to LDA: $2.58$ eV, full GW approximation: $2.19$ eV).
However, in the non-degenerate multi-band materials such as FeSe and FeTe, the momentum dependent self-energy effects yield substantial modifications of the band structures and relative shifts of orbital-energy levels of the effective models.  
The FeSe model indicates a substantial downward shift of the $x^2-y^2$ orbital level, which may lead to an increase in the filling of the $x^2-y^2$ orbital above half filling.
It suggests a destruction of the orbital selective Mott phase of the $x^2-y^2$ orbital in agreement with the experimental absence of the antiferromagnetic phase.
\end{abstract}


\maketitle


\section{Introduction}

Strongly correlated electron systems have been a central subject of condensed matter research for decades. 
A challenge is to unveil their electronic structures from first principles. 
A major advance has recently been made in this direction~\cite{imada,kotliar06} by taking advantage of the hierarchical energy structure of strongly-correlated-electron materials, which commonly have only a small number of bands near the Fermi level in contrast to usual highly dense band entanglement at energies far away from the Fermi level. In this general framework of multiscale \textit{ab initio} scheme for correlated electrons (MACE)~\cite{imada}, global dense and entangled band structure is calculated based on density functional theory (DFT)~\cite{hohenberg,kohn}. The degrees of freedom far away from the Fermi level is traced out perturbatively in the spirit of the renormalization group, which generates an effective low-energy model consisting of a small number of bands near the Fermi level, which we call ``target bands". This downfolding procedure is efficiently performed by the constrained random phase approximation (cRPA)~\cite{aryasetiawan1,solovyev1} after constructions of the maximally localized Wannier orbitals~\cite{marzari,souza}. Derived {\it ab initio} low-energy models are represented in general by multi-band fermions on lattices, which are typically extended-Hubbard-type models. The low-energy models are solved by high-accuracy solvers such as variational Monte Carlo (VMC)\cite{tahara}, path-integral renormalization group (PIRG)\cite{imada2, mizusaki}, dynamical mean field theory (DMFT)\cite{metzner, georges} with its cluster extensions~\cite{maier05,kotliar01,kotliar06,Senechal,Potthoff1,Potthoff2,Rubtsov08,Rubtsov09,ToschiHeld07,ToschiHeld08,Sangiovanni10} and functional renormalization group (FRG)~\cite{metzner12}.
The validity of the cRPA is ascribed to the small vertex correction thanks to the above hierarchical structure~\cite{imada}, while an extension of the cRPA by using the FRG has also been proposed~\cite{honerkamp}.

This hybrid scheme referred inclusively as MACE has been applied to a wide variety of materials; semiconductors\cite{nakamura1}, transition metals\cite{solovyev1, solovyev2, miyake,miyake09}, transition-metal-oxides~\cite{aryasetiawan06,miyake, pavarini, imai1, imai2, otsuka,biermann05,tomczak,tomczak2,weber,casula}, transition-metal-oxide interface\cite{hirayama}, molecular organic conductors\cite{nakamura4,shinaoka,kandpal,jeschke}, 
and Fe-based layered superconductors\cite{nakamura2, miyake2} by using the solvers with the choices of VMC~\cite{misawa11,misawa12}, DMFT~\cite{liebsch,yin-haule,ferber,aichhorn1,aichhorn2,werner} and the FRG\cite{platt,thomale}. 
{\it Ab initio} effective models were derived for other challenging complex materials with a large number of atoms on a unit cell as well~\cite{nomura,nohara}.

Quantitative accuracies of obtained physical quantities in comparison to experimental results support the validity of the scheme. However, there remains a basic question about the counting of the electron correlation: The correlation effects are, though insufficiently, taken into account, in principle, in the DFT calculation as the exchange correlation energy.
Since the correlation effects contained in the effective low-energy model are considered when the model is solved by the low-energy solver, there exists a well-known double counting problem.
Removing the Hartree contribution can be considered easily when solved by the low-energy solver\cite{misawa11}. On the other hand, the double counting arising from the exchange correlation has simply been assumed to be small in most of the studies in the literature.
Removing the double counting as well as removing the influence of the original LDA/DFT bias is a desired step for the consistent and {\it ab initio} descriptions of the correlation effects in the present downfolding scheme. Several attempt in this direction has been put forward in the literature~\cite{biermann03,karolak,wang,nekrasov}. 
Since it is not a practical choice to completely throw away the DFT/LDA calculations in obtaining the global band structure, it is better to remove the bias and the double counting as a correction as much as possible within the present framework.

If the correlation effects are taken into account partially in the exchange correlation in the LDA calculation, narrowing of the bands coming from the effects within the target bands, however small it is, is regarded as the double counting, and the resultant parameters underestimate the transfer integral in the effective model. 
   
To overcome the double-counting problem, in the first step of this paper, we subtract the whole exchange correlation energy in the LDA to remove the double counting and to completely eliminate the bias of the LDA.  Instead of the exchange-correlation potential $V^{\text{xc}}$, we then add a constrained self-energy arising from the high-energy region $\Sigma ^{H}=-G^{H}W$ based on the GW approximation (GWA)~\cite{hedin,aryasetiawan2} by excluding the self-energy contribution from the low-energy target degrees of freedom. Here, $G^{H}$ is the Green's function for the degrees of freedom outside the target bands (high-energy bands) and $W$ is the fully screened interaction in the RPA as we define below.

In general, the one-body parameter calculated from such constrained GWA and the two-body parameter calculated from the cRPA have frequency dependences originating from the constrained self-energy $\Sigma^{H}(\omega)$ and the constrained polarization $P^{H}(\omega)$, respectively. 
Although the frequency dependence of the interaction has been considered in the literature~\cite{casula,casula2,werner}, in practice, it is difficult to treat the dynamical Hubbard model by keeping both the frequency and momentum dependences of fluctuations at high accuracy.
So far, although DMFT is able to incorporate frequency dependences of the effective interaction, it is actually difficult to consider spatial fluctuations accurately in DMFT.
Therefore, in this paper we make an attempt to derive reliable {\it static} effective fermion models on lattices. 
 
Then, in the second step of the present study, we renormalize the frequency dependence of the effective interaction to derive static low-energy effective models.
This Hamiltonian formalism allows the consideration of spatial fluctuation effects accurately by low-energy solvers.
We renormalize the frequency dependence of the constrained self-energy $\Sigma^{H}$ into static one-body parameters by using the renormalization factor $Z^{H}$, namely by using the linear frequency dependence of the real part of $\Sigma^{H}$.
As we will see later, the constrained self-energy has a well-behaved and gentle frequency dependence near the Fermi level, and thus the expansion around the energy eigenvalue is justified.
In addition, we renormalize the effect of frequency dependence of the partial screened Coulomb interaction $W^{p}(\omega)$ into the one-body part as the self-energy~\cite{aryasetiawan1}. 

We next apply the present formalism to several typical examples: First we apply to SrVO$_3$, where the $t_{2g}$ orbital degeneracy is retained. Second we apply to FeSe and FeTe, typical iron-based superconductors, where the lift of the orbital degeneracy introduces additional complexity. 

In these examples, we show that after removing the double counting in the low-energy space obtained in the first step, the bandwidth and the transfer integral increase typically about $30$-$50$\% than those of the LDA. 
When we perform the second step and renormalize the effect of the frequency dependence of the effective Coulomb interaction screened by the high-energy bands into the mass enhancement, then the bandwidth and the transfer integral remarkably recovers to those of the LDA. 

However, in the case of the multi-band systems with non-degenerate orbitals, the removal of the double counting causes the orbital dependent shift of the chemical potential.
Because the constrained self-energy is not only frequency dependent but also wavenumber and orbital dependent, the low-energy band after excluding the double-counting is different from that of the LDA, especially in non-equivalent multi-orbital systems such as Fe-based layered superconductors. However, A warning is that this band structure should not be compared with experiments, because the experimental band structure is a consequence of the many-body effects within the target band and the direct comparison is possible only after solving the low-energy effective model.  

In Sec.2 we propose the formulation. 
Section 3 describes the global band structures together with the derived effective models after the improved downfolding for the examples of a transition-metal-oxide SrVO$_{3}$ (SVO) and the  Fe-based layered superconductors, FeSe and FeTe. 
Section 4 is devoted to discussions and summary. 
 

\section{Method}

\subsection{Constrained Random Phase Approximation}

In the LDA, the band structure is calculated from the Kohn-Sham equation
\begin{multline}
\mathcal{H}^{\text{LDA}}\psi_{i} ^{\text{LDA}}(r) \\
= [-\frac{1}{2}\nabla ^2 + V^{\text{ext}}(r)+ V^{\text{H}}(r)+V^{\text{xc}}(r)]\psi_{i} ^{\text{LDA}}(r) \\
=\epsilon _{i} ^{\text{LDA}} \psi_{i} ^{\text{LDA}}(r),
\label{LDAeq}
\end{multline}
where  $V^{\text{ext}}$ is the potential of atomic nucleus, $V^{\text{H}}$ is the Hartree term of the Coulomb interaction, and $V^{\text{xc}}$ is the exchange-correlation potential. 
The energy eigenvalue is denoted by $\epsilon _{i} ^{\text{LDA}} $, and $\psi_{i} ^{\text{LDA}}$ is its eigenstate. 

Now we derive the effective model for the target bands as a typical case. Our target bands here are the $d$ electrons. 
From the eigenvalues and eigenstates of Eq. (\ref{LDAeq}), the transfer integral of the low-energy effective model is given by 
\begin{equation}
t^{\text{LDA}}_{mn}(\bm{R})= \langle \phi^{L}_{m\bm{0}}|\mathcal{H}^{\text{LDA}}|\phi^{L}_{n\bm{R}} \rangle,
\label{tLDA}
\end{equation}
where $\phi^{L}_{m\bm{R}}$ is the maximally localized Wannier function (MLWF) of the $m$th orbital localized at the unit cell $\bm{R}$ in the low-energy space derived from a linear combination of the Kohn-Sham (KS) wave functions by following the conventional prescription\cite{marzari,souza}. 

The partially screened Coulomb interaction for the $d$ bands is given by
\begin{equation}
U_{mn}(\bm{R},\omega )= \langle \phi^{L}_{m\bm{0}}\phi^{L}_{m\bm{0}}|W^{p}|\phi^{L}_{n\bm{R}}\phi^{L}_{n\bm{R}}\rangle ,
\label{U}
\end{equation}
\begin{equation}
W^{p}=\frac{v}{1-vP^{H}},
\label{Wp}
\end{equation}
where $v$ is the bare Coulomb interaction, and $P^{H}$ is the polarization without low-energy $L$-$L$ transition, $P^{H}=P-P^{L}$. Namely, $P^H$ is defined from the total polarization subtracted from $P^L$, the polarization contributed only from the $L$ space.  
The indices ``$L$" and ``$H$" express the low- (within target bands) and high-energy (including outside of the target bands) parts of the Hilbert spaces, respectively.\cite{aryasetiawan1}
The frequency dependence of $W^{p}$ comes from the frequency dependence of the high-energy polarization $P^H$. 
The partially screened Coulomb interaction is usually written as $W_r$ in the literature, but, in this paper, we express it as $W^{p}$ to avoid confusions with the renormalized Green's function in the low-energy space.


\subsection{Double-Counting of Self-Energy in Low-Energy Space}
The transfer integral in Eq. (\ref{tLDA}) formally contains the self-energy of the low-energy space in the form of the exchange-correlation energy $V^{\text{xc}}$ in the DFT/LDA.   
However, the parameters of the low-energy effective model should not contain such an energy of the low-energy space, 
because the self-energy effect in the low-energy space should be considered when one solves the
low-energy effective models.
Therefore, this contribution has to be subtracted, but it is not easy to separate {$V^{\text{xc}}(n(\bm{r})_{\rm all})$ to the high- and low-energy parts.

On the other hand, in the GWA beyond the LDA, we can easily divide the whole self-energy $GW$ into the low- and high-energy parts:
\begin{equation}
GW= G^{L}W+G^{H}W,
\end{equation}
where the whole Green's function $G$ is given by the sum of the low- and high-energy propagators, $G^L$ and $G^H$, respectively, and $W$ is the fully screened Coulomb interaction.

When one subtracts the exchange-correlation energy $V^{\text{xc}}$ from the one-body energy in the Kohn-Sham LDA Hamiltonian $\mathcal{H}^{\text{LDA}}$ and add a constrained self-energy  $\Sigma^{H}=-G^{H}W$ to that, we can calculate the transfer integral by excluding the self-energy contributed from the low-energy space only:  
\begin{equation}
\tilde{t}^{H}_{mn}(\bm{R},\omega )= \langle \phi^{L}_{m\bm{0}}|\mathcal{H}^{\text{LDA}}-V^{\text{xc}}+\Sigma^{H}|\phi^{L}_{n\bm{R}} \rangle .
\label{tGrWomega}
\end{equation}
This allows to eliminate the double counting of the self-energy in the low-energy space when 
one solves the low-energy model honestly.

Moreover, it allows a formalism to calculate the one-body parameter beyond the DFT/LDA schemes.
The good convergence of the GW scheme in terms of the perturbative expansion and the irrelevance of the vertex correction\cite{imada} assure the accuracy of $\Sigma^H$.

\subsection{Static Hubbard Model}

In the previous section, we consider the ``dynamical" effective Hubbard model
\begin{multline}
\mathcal{H} ^{\text{eff}}= \sum_{ij} \sum_{mn\sigma }\tilde{t}^{H}_{mn\sigma}(\bm{R}_i-\bm{R}_j, \omega ) d_{in\sigma} ^{\dagger} d_{jm\sigma} \\
+ \frac{1}{2} \sum_{ij} \sum_{mn \sigma \rho} \biggl\{ U_{mn\sigma \rho }(\bm{R}_i-\bm{R}_j, \omega ) d_{in\sigma}^{\dagger}  d_{jm\rho}^{\dagger} d_{jm\rho} d_{in\sigma} \\ 
+J_{mn\sigma \rho}(\bm{R}_i-\bm{R}_j, \omega ) \bigl( d_{in\sigma}^{\dagger} d_{jm\rho}^{\dagger} d_{in\rho} d_{jm\sigma} \\
+d_{in\sigma}^{\dagger} d_{in\rho}^{\dagger} d_{jm\rho} d_{jm\sigma}\bigr) \biggr\}, 
\label{eqH}  
\end{multline} 
where $d_{in\sigma} ^{\dagger}$ ($d_{in\sigma}$) is a creation (annihilation) operator of an electron with spin $\sigma$ in the $n$th MLWF centered at $\bm{R}_{i}$. The effective Coulomb interaction $U$ and the exchange interaction $J$ are given by
\begin{equation}
U_{mn}(\bm{R}) =  \langle \phi^{L}_{m\bm{0}}\phi^{L}_{m\bm{0}}|W^{p}|\phi^{L}_{n\bm{R}}\phi^{L}_{n\bm{R}}\rangle ,
\end{equation} 
and 
\begin{equation}
J_{mn}(\bm{R}) =  \langle \phi^{L}_{m\bm{0}}\phi^{L}_{n\bm{0}}|W^{p}|\phi^{L}_{n\bm{R}}\phi^{L}_{m\bm{R}}\rangle ,
\end{equation} 
respectively.

The renormalized Green's function of the low-energy space $G^{Lp}$, where the constrained self-energy from the high-energy space is implemented, is written as 
\begin{align}
&G^{Lp}(\omega ) =\frac{G^{L0}}{1-G^{L0}\Sigma^{H}} \\
&= \frac{1}{\omega -(H^{\text{LDA}}-V^{\text{xc}}+\Sigma^{H}(\omega ))} \\
\approx &\frac{1}{\omega -(H^{\text{LDA}}-V^{\text{xc}}+\Sigma^{H}(\epsilon ^{\text{LDA}})+\frac{\partial \Sigma^{H}}{\partial \omega}|_{\omega =\epsilon ^{\text{LDA}}}(\omega -\epsilon ^{\text{LDA}}) )} \label{rGexp} \\
              &= \frac{Z^{H}(\epsilon ^{\text{LDA}})}{\omega -(H^{\text{LDA}}+Z^{H}(\epsilon ^{\text{LDA}})(-V^{\text{xc}}+\Sigma^{H}(\epsilon ^{\text{LDA}})))} ,
\label{rG}
\end{align}
where the ``bare" Green's function $G^{L0}$ is
\begin{eqnarray}
G^{L0}(\omega )
              &=&\frac{1}{\omega -(H^{\text{LDA}}-V^{\text{xc}})},
\label{dG}
\end{eqnarray}
and
$Z^{H}(\epsilon ^{\text{LDA}})$ is the renormalization factor of $\Sigma ^{H}$: 
\begin{equation}
Z^{H}(\epsilon )= \biggl\{ 1-\frac{\partial \text{Re}\Sigma ^{H}}{\partial \omega }\Big|_{\omega =\epsilon } \biggr\}^{-1}.
\end{equation}
We note that $H^{\text{LDA}}$ is not renormalized by $Z^H$, because
the Fermi level must be fixed by the electron number and we have taken into account the $\omega$ dependence around $\epsilon^{\text{LDA}}$.
Because $\Sigma^{H}=-G^{H}W$ does not contain the poles of $G^{L}$, $Z^{H}$ is close to $1$ in contrast to the substantially small renormalization factor obtained for the full self-energy $GW$, as we will see later. 

If we can ignore the imaginary part of $\Sigma^{H}$ and $Z^{H}$ in the numerator of $G^{Lp}(\omega ) $, and the terms higher than the linear one in the $\omega$ dependence in the real part of $\Sigma^{H}$, we obtain
\begin{multline}
G^{Lp}(\omega )= \\
 \frac{1}{\omega -(H^{\text{LDA}}+Z^{H}(\epsilon ^{\text{LDA}})(-V^{\text{xc}}+\text{Re}\Sigma^{H}(\epsilon ^{\text{LDA}})))} ,
\label{rG2}
\end{multline} 
or, in the case where the self-energy is expanded around the Fermi level instead of the eigen value in the LDA,
\begin{equation}
 G^{Lp}(\omega ) =  \frac{1}{\omega -Z^{H}(0)(H^{\text{LDA}}-V^{\text{xc}}+\text{Re}\Sigma^{H}(0))},
\label{rG20}
\end{equation}
which preserves the number of the electron in eq. (\ref{rG}) and has the form of the ``dynamical" effective Hubbard model, containing the $\omega $ dependence in the two-body part,

\begin{align}
\mathcal{H} ^{\text{eff}}&= \sum_{ij} \sum_{mn\sigma } t^{H}_{mn\sigma}(\bm{R}_i-\bm{R}_j) d_{in\sigma} ^{\dagger} d_{jm\sigma} \nonumber \\
+ \frac{1}{2}& \sum_{ij} \sum_{mn \sigma \rho} \biggl\{ U_{mn\sigma \rho }(\bm{R}_i-\bm{R}_j, \omega ) d_{in\sigma}^{\dagger}  d_{jm\rho}^{\dagger} d_{jm\rho} d_{in\sigma} \nonumber \\ 
&+J_{mn\sigma \rho}(\bm{R}_i-\bm{R}_j, \omega ) \bigl( d_{in\sigma}^{\dagger} d_{jm\rho}^{\dagger} d_{in\rho} d_{jm\sigma} \nonumber \\
&\ \ \ \ \ \ \ \ \ \ \ \ \ \ \ \ \ \ \ \ \ \ \ \ \ 
+d_{in\sigma}^{\dagger} d_{in\rho}^{\dagger} d_{jm\rho} d_{jm\sigma}\bigr) \biggr\}, 
\label{eqHz}  
\end{align}
where the static one-body part $t^{H}$ is given by
\begin{equation}
t^{H}_{mn}(\bm{R})= \langle \phi^{L}_{m\bm{0}}|\mathcal{H}^{H}|\phi^{L}_{n\bm{R}} \rangle, 
\label{tGrW}
\end{equation}
\begin{multline}
\mathcal{H}^{H}= \mathcal{H}^{\text{LDA}}+Z^{H}(\epsilon ^{\text{LDA}})(-V^{\text{xc}}+\text{Re}\Sigma^{H}(\epsilon ^{\text{LDA}})) \\
= Z^{H}(0)(\mathcal{H}^{\text{LDA}}-V^{\text{xc}}+\text{Re}\Sigma^{H}(0)) .
\label{HGrW}
\end{multline}
In Eq. (\ref{HGrW}), we employed the renormalization factor $Z^{H}$ only with multiplying by the diagonal part of Re$\Sigma^H$, whereas $Z^H$ for the off-diagonal elements is fixed to unity. 
Eigenstates and $\mathcal{H}^{H}$ are determined self-consistently.

Actually, the $\omega$ dependence requires the Lagrangian description instead of the Hamiltonian description.  However, if the frequency dependence is small in the energy range of physics to be considered by the low-energy effective model, one may ignore the frequency dependence by replacing $U(\omega)$ and $J(\omega)$ with its static limit $\omega \rightarrow 0$. This is indeed the case when the low-energy bands are well separated from the high-energy ones
as is normally expected in the strongly correlated electron systems~\cite{imada}.  
Then one may employ a static value
\begin{equation}
W^{p}(\omega=0)= \frac{v}{1-vP^{H}(0)}
\end{equation}
instead of Eq. (\ref{Wp}).

To further renormalize the effects of $\omega $-dependence of $P^{H}$ in the GWA, we append the self-energy correction. A choice of this correction is
\begin{equation}
-\Delta \Sigma ^{L}_{\text{HF}}=G^{Lp}(W^{p}(\omega )-W^{p}(0) ) . 
\label{dsig}
\end{equation}
Note that $W^{p}(\omega)-W^{p}(0)$ vanishes at $\omega=0$.
Equation (\ref{dsig}) is the ``Hartree-Fock" approximation (HFA), because $G^{Lp}$ is the ``bare" Green's function in the sense of the absence of the self-energy arising from $L$-electron's many-body effects and $W^{p}$ is the ``bare" Coulomb interaction in terms of the low-energy effective model. 

In the case of the GWA beyond the HFA, the self-energy correction is calculated as
\begin{equation}
-\Delta \Sigma ^{L} _{\text{GW}}=G^{L}(W-W^{U}) , 
\label{dsiggw}
\end{equation}  
where 
\begin{equation}
W^{U}= \frac{W^{p}(0)}{1-W^{p}(0)P^{L}(\omega)}.
\end{equation}
Here $W^{U}(\omega)$ is equal to the fully screened interaction $W(\omega)$ at $\omega=0$, while 
$W^{U}(\omega)$ approaches $W^p(0)$ for $\omega \rightarrow \infty$.
On the other hand, $W$ approaches $v$ in the $\omega \rightarrow \infty$ limit.  
In this one-shot GWA, the self-energy expected in the level of the GWA of the effective model,
namely $G^LW^{U}$,
has been subtracted from the full GW self-energy $G^LW$, since $G^LW^{U}$ should be considered when one solves the model.
In this paper, we calculate $\Delta \Sigma ^{L}$ by the one-shot GWA according to Eq. (\ref{dsiggw}).

Thus, the renormalized Green's function for the static Hubbard model   
\begin{align} 
&G^{Lp\Delta } (\omega )=\frac{G^{Lp}}{1-G^{Lp}\Delta \Sigma ^{L}} \\
& \ \ \ \ \ \ \ \ \ \ \ =\frac{1}{\omega -(H^{\text{LDA}}-V^{\text{xc}}+\Sigma ^{H}+\Delta \Sigma ^{L})}  \\
                       \approx &\frac{Z^{H\Delta}(\epsilon ^{\text{LDA}})}{\omega -(H^{\text{LDA}}+Z^{H\Delta}(\epsilon ^{\text{LDA}})(-V^{\text{xc}}+(\Sigma ^{H}+\Delta \Sigma ^{L})(\epsilon ^{\text{LDA}})))}  ,
\label{rdelG}
\end{align}
\begin{equation}
Z^{H\Delta}(\epsilon)= \biggl\{ 1-\frac{\partial (\text{Re}\Sigma ^{H}+\text{Re}\Delta \Sigma ^{L})}{\partial \omega }\Big|_{\omega =\epsilon } \biggr\}^{-1} .
\label{Zrdel}
\end{equation}
is obtained.

Similarly to the reduction to Eq. (\ref{rG2}), even with this $\Delta\Sigma^L$ correction as well, we reduce the renormalized Green's function with $Z$ factor to
\begin{multline} 
G^{Lp\Delta } (\omega )= \\
\frac{1}{\omega -(H^{\text{LDA}}+Z^{H\Delta}(\epsilon ^{\text{LDA}})(-V^{\text{xc}}+\text{Re}(\Sigma ^{H}+\Delta \Sigma ^{L})(\epsilon ^{\text{LDA}})))} \\
=\frac{1}{\omega -Z^{H\Delta}(0)(H^{\text{LDA}}-V^{\text{xc}}+\text{Re}(\Sigma ^{H}+\Delta \Sigma ^{L})(0))} .
\label{rdelG2}
\end{multline}
In the practical calculation below, we employ $\Delta \Sigma ^{L}_{\text{HF}}$ in Eq. (\ref{dsig}) for $\Delta \Sigma ^{L}$.
Then the ``static" effective Hubbard model, where all the effects of frequency dependences in the one- and two-body parts are incorporated into the renormalization of the one-body part as
\begin{multline}
\mathcal{H} ^{\text{eff}}= \sum_{ij} \sum_{mn\sigma } t^{H\Delta}_{mn\sigma}(\bm{R}_i-\bm{R}_j) d_{in\sigma} ^{\dagger} d_{jm\sigma} \\
+ \frac{1}{2} \sum_{ij} \sum_{mn \sigma \rho} \biggl\{ U_{mn\sigma \rho }(\bm{R}_i-\bm{R}_j, 0) d_{in\sigma}^{\dagger}  d_{jm\rho}^{\dagger} d_{jm\rho} d_{in\sigma} \\ 
+J_{mn\sigma \rho}(\bm{R}_i-\bm{R}_j, 0) \bigl( d_{in\sigma}^{\dagger} d_{jm\rho}^{\dagger} d_{in\rho} d_{jm\sigma} \\
+d_{in\sigma}^{\dagger} d_{in\rho}^{\dagger} d_{jm\rho} d_{jm\sigma}\bigr) \biggr\}, 
\label{eqHz2}  
\end{multline} 
where the renormalized static one-body part $t^{H\Delta}$ is given by 
\begin{equation}
t^{H\Delta}_{mn}(\bm{R})= \langle \phi^{L}_{m\bm{0}}|\mathcal{H}^{H\Delta}|\phi^{L}_{n\bm{R}} \rangle, 
\label{tGrWdelta}
\end{equation}
\begin{multline}
\mathcal{H}^{H\Delta}= \\
\mathcal{H}^{\text{LDA}}+Z^{H\Delta}(\epsilon ^{\text{LDA}})(-V^{\text{xc}}+\text{Re}(\Sigma^{H}+\Delta \Sigma ^{L})(\epsilon ^{\text{LDA}})) \\
=Z^{H\Delta}(0)(\mathcal{H}^{\text{LDA}}-V^{\text{xc}}+\text{Re}(\Sigma^{H}+\Delta \Sigma ^{L})(0)) .
\label{HGrWdelta}
\end{multline}

The renormalized effective models of Eqs. (\ref{HGrW}) and (\ref{HGrWdelta}) with Eqs. (\ref{eqHz}) and (\ref{tGrWdelta}) still contain a double counting of the Hartree term arising from the low-energy (target) degrees of freedom in the one-body part.
This Hartree double counting may easily be subtracted in the model\cite{misawa11}, because the Hartree term in the LDA and that in the model are essentially the same. 
For example, to subtract the double counting of the Hartree term for the on-site potential, one should employ the one-body part as  
\begin{equation}
t^{\text{dcf}}_{mm\sigma}(\bm{R}_i)= t^{H\Delta}_{mm\sigma}(\bm{R}_i) 
-\sum _{jn\rho } U_{mn\sigma \rho }(\bm{R}_i-\bm{R}_j,0) \langle n^{L}_{jn\rho} \rangle,
\label{muHap}
\end{equation}
where the average $\langle n^{L}_{jn\rho} \rangle$ is calculated from the Hartree approximation of $\mathcal{H}^{\text{LDA}}$ combined with the partially screened interaction calculated from cRPA as listed in Ref. \cite{miyake2}. 
This prescription is based on the observation that the Hartree potential originated from the low-energy target degrees of freedom considered in the LDA may well be reproduced in the Hartree approximation of the previous model obtained without considering the self-energy effect, because the Hartree potential must be essentially the same.
Then we can determine the on-site potential $t^{\text{dcf}} _{mm\sigma}(\bm{R}_i)$ free from the double counting.

\subsection{Computational Conditions}
Computational conditions are as follows.
We calculate the band structures of the transition-metal-oxide SVO, and the Fe-based layered superconductors FeSe and FeTe based on the DFT/LDA\cite{hohenberg,kohn}.
The band structure calculation is based on the full-potential linear muffin-tin orbitals (FP-LMTO) implementation\cite{methfessel}.
The cRPA and GW calculations use a mixed basis consisting of products of two atomic orbitals and interstitial plane
waves~\cite{schilfgaarde06}.
In the LDA calculation, $8\times 8\times 8$ $k$-mesh is employed for the SVO, $12\times 12\times 6$ $k$-mesh is employed for FeSe and FeTe.
In the cRPA and GW calculation, $6\times 6\times 6$ $k$-mesh is employed for the SVO, $3\times 3\times 3$ $k$-mesh is employed for FeSe and FeTe.


\section{Result}


\subsection{SrVO$_{3}$}

\begin{figure}[ptb]
\centering 
\includegraphics[clip,width=0.4\textwidth ]{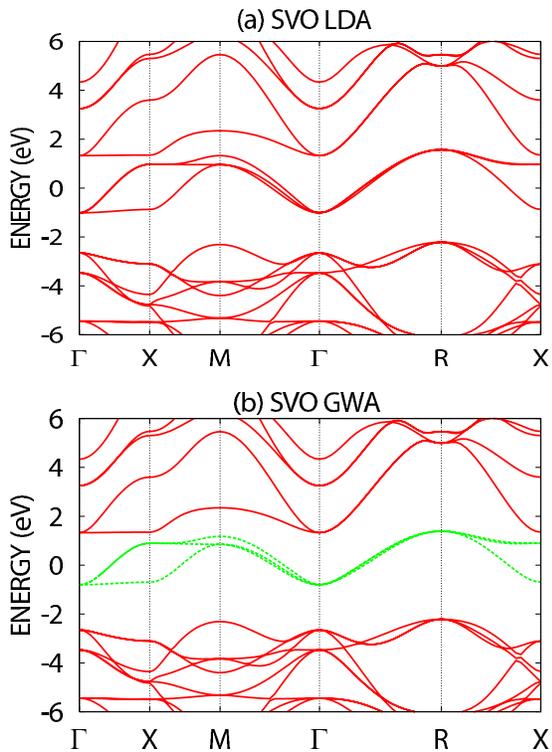} 
\caption{(Color online) Electronic band structures of SVO obtained by (a) the LDA and (b) the GWA. The zero energy corresponds to the Fermi level. In the GWA, the self-energy is calculated only for the $t_{2g}$ Wannier band.}
\label{bndsSVOLDAandGW}
\end{figure} 
\begin{figure}[ptb]
\centering 
\includegraphics[clip,width=0.2\textwidth ]{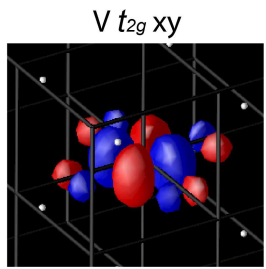} 
\caption{(Color online) Isosurface of the maximally localized Wannier function for $\pm 0.05$ a.u. for the V $d_{xy}$ orbitals in SVO. 
}
\label{WannierSVO}
\end{figure} 
\begin{table}[ptb] 
\caption{Band width of the $t_{2g}$ band in each calculation of the self-energy. Units are given in eV. 
The band width of "$\mathcal{H}^{H}$" is calculated according to Eq. (\ref{rG2}) and that of $\mathcal{H}^{H\Delta}$ is calculated according to Eq. (\ref{rdelG2}) respectively.
The band width except for "LDA" includes the $\omega$-dependence effect of the self-energy through the renormalization factor.  
} 

\begin{tabular}{c|cccc}
\hline \hline \\ [-8pt]  
SVO   &  LDA  &  GWA & $\mathcal{H}^{H}$ & $\mathcal{H}^{H\Delta}$   \\ [+1pt]
\hline \\ [-8pt] 
width &  2.58 &  2.19 & 3.41 & 2.56    \\ 
\hline \hline 
\end{tabular}
\label{width_SVO} 
\end{table}

\begin{figure*}[ptb]
\begin{center} 
\includegraphics[width=0.95\textwidth ]{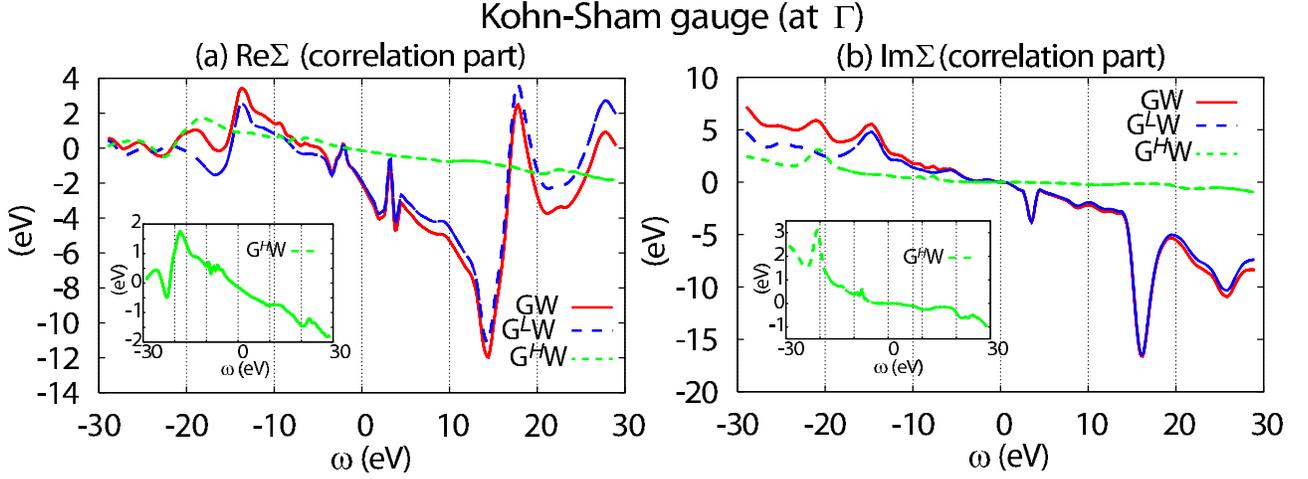} 
\end{center} 
\caption{(Color online) (a) Real parts and (b) imaginary parts of correlation self-energy of the $t_{2g}$ Bloch functions at $\Gamma=(0,0,0)$ in the Kohn-Sham gauge. The zero energy corresponds to the Fermi level. Inner window is the magnification only for "G$^{H}$W".}
\label{KSSigmaSVO}
\end{figure*} 

In this section, we derive the low-energy effective model for the V $t_{2g}$ bands of the transition-metal-oxide SVO.
Figure~\ref{bndsSVOLDAandGW} (a) shows the band structure of SVO in the DFT/LDA.
SVO is a paramagnetic metal\cite{onoda} and has a cubic structure with lattice parameter $a_{\text{SVO}}=3.843$ \AA \cite{lan}.  
The three conduction bands at the Fermi level are derived from the $t_{2g}$ orbitals of V sites, where the octahedral crystal field of O$^{-2}$ partially breaks the five-fold symmetry of the $3d$ orbitals into the lower orbitals of the $t_{2g}$ and the higher orbitals of the $e_{g}$.
The band width of the $t_{2g}$ bands is $2.58$ eV.
One conduction electron per unit cell is accommodated in the $t_{2g}$ bands. 
Hereafter, we regard such three conduction bands of the $t_{2g}$ orbitals as the low-energy bands and 
the rest of the whole band structure of SVO as the high-energy bands. 
The top of the occupied high-energy band is the O $2p$ band and the bottom of the unoccupied high-energy band is the V $e_{g}$ band. 

\begin{figure*}[ptb]
\begin{center} 
\includegraphics[width=0.85\textwidth ]{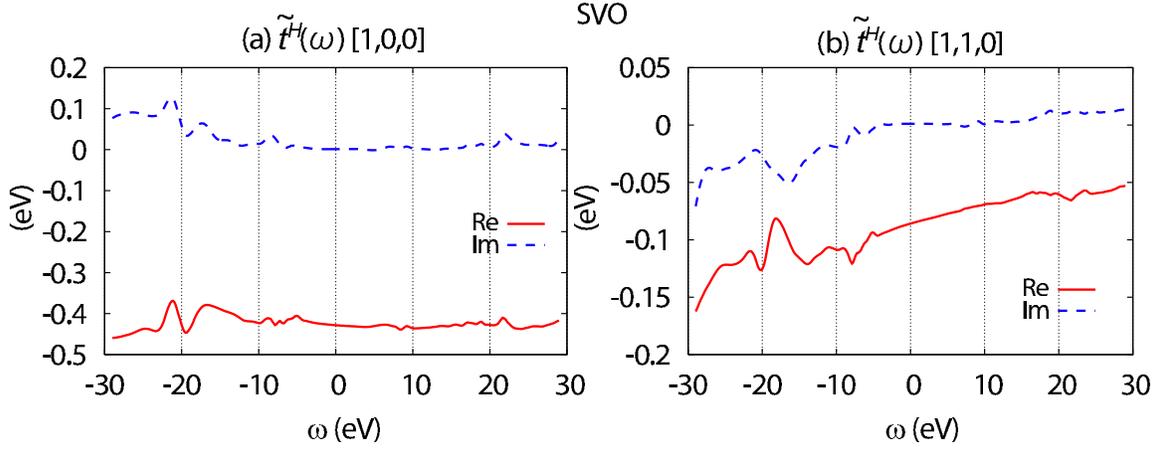} 
\end{center} 
\caption{(Color online) Dynamical transfer integral of the $t_{2g}$ orbitals of the V sites without double-counting terms: $\tilde{t}^{H}(\omega )= \langle \phi^{L}|\mathcal{H}^{\text{LDA}}-V^{\text{xc}}+\Sigma^{H}|\phi^{L} \rangle$ (see Eq. (\ref{tGrWomega})).
The left panel is the nearest hopping $[1,0,0]$ of the V $t_{2g}$ orbitals and the right panel is the next-nearest hopping $[1,1,0]$.
These values are dynamical and, therefore, do not include the effect of the renormalization factor $Z^{H}$.
}
\label{tGrWomegaSVO100and110}
\end{figure*} 

Figure~\ref{bndsSVOLDAandGW} (b) shows the band structure of SVO in the GWA, where we calculate the self-energy only for the V $t_{2g}$ orbitals after removing the exchange correlation energy in the LDA.
Figure~\ref{WannierSVO} shows the isosurface of the Wannier function of the V $d_{xy}$ orbital, where the V $d_{xy}$ atomic orbital hybridizes the nearest O $p$ atomic orbitals.
The Fermi energy of the low-energy bands in Fig.~\ref{bndsSVOLDAandGW} (b) is determined by the occupation number in the low-energy space separated from the whole Hilbert space.     
Due to the large self-energy mainly from the low-energy space of the $t_{2g}$ bands, the band width in the GWA is about $15$\% smaller than that in the LDA (see Table \ref{width_SVO}).
In Fig.~\ref{KSSigmaSVO}, the $\omega$-dependence of the self-energy of SVO in the one-shot GWA is not smooth near the eigenvalues of the V $t_{2g}$ bands.
Incidentally, when one chooses [$-10, 10$] (eV) instead of [$-0.5, 0.5$] (eV) as the $\omega$ range to calculate the renormalization factor, then the average of $Z$ considerably increases from $0.55$ to $0.74$.  
This is an indication that the linear approximation of the self-energy collapses in the low-energy band near the Fermi level.

\begin{table*}[htb] 
\caption{Renormalization factor $Z$ for each V $t_{2g}$ Wannier band in bulk SVO. $Z^{H}$ is the partial renormalization factor: $Z^{H}$= $(1-\frac{\partial \Sigma ^{H}}{\partial \omega}|_{\omega =0})^{-1}$ . "Average" is the average of the renormalization factor in $k$-space. Each renormalization factor calculated from the energy-range [-0.5,0.5] (eV) around the Fermi level except for that with index "wide" which is calculated from the energy-range [-10,10] (eV) around the Fermi level.} 
\
\begin{tabular}{c|ccccc|c|ccccc}
\hline \hline \\ [-8pt]   
$Z$  & Average & $\Gamma$ &    X &    M & R    & $Z_{\text{wide}}$ & Average & $\Gamma$ &    X &    M &    R  \\ 
\hline \\ [-8pt] 
$xy$ &    0.55 &     0.54 & 0.53 & 0.56 & 0.57 &              $xy$ &    0.74 &     0.75 & 0.75 & 0.73 & 0.73  \\ 
$yz$ &    0.55 &     0.54 & 0.56 & 0.56 & 0.57 &              $yz$ &    0.74 &     0.75 & 0.73 & 0.73 & 0.73  \\
$zx$ &    0.55 &     0.54 & 0.56 & 0.57 & 0.57 &              $zx$ &    0.74 &     0.75 & 0.73 & 0.73 & 0.73  \\
\hline \\ [-8pt]
$Z^{L}$ & Average & $\Gamma$ &    X &    M &    R & $Z^{H}$ & Average & $\Gamma$ &    X &    M &    R  \\ 
\hline \\ [-8pt] 
$xy$    &    0.58 &     0.56 & 0.55 & 0.59 & 0.60 & $xy$    &    0.92 &     0.93 & 0.93 & 0.92 & 0.93  \\ 
$yz$    &    0.58 &     0.56 & 0.59 & 0.59 & 0.60 & $yz$    &    0.92 &     0.93 & 0.91 & 0.92 & 0.93  \\
$zx$    &    0.58 &     0.56 & 0.59 & 0.59 & 0.60 & $zx$    &    0.92 &     0.93 & 0.91 & 0.93 & 0.93  \\
\hline \\ [-8pt] 
 $Z^{\Delta}$ & Average & $\Gamma$ &  X &  M & R & $Z^{H\Delta}$ & Average & $\Gamma$ &  X &  M & R   \\ 
\hline \\ [-8pt] 
 $xy$    &    0.77 &     0.78 & 0.77 & 0.78 & 0.77 & $xy$    &    0.72 &     0.73 & 0.73 & 0.72 & 0.73    \\ 
 $yz$    &    0.77 &     0.78 & 0.78 & 0.78 & 0.77 & $yz$    &    0.72 &     0.73 & 0.73 & 0.72 & 0.73  \\
 $zx$    &    0.77 &     0.78 & 0.78 & 0.77 & 0.77 & $zx$    &    0.72 &     0.73 & 0.73 & 0.73 & 0.73  \\
\hline \hline 
\end{tabular}
\label{Z_SVO} 
\end{table*}

To calculate the transfer integral without double-counting terms in the low-energy space, we calculate the constrained self-energy originated only from the high-energy space, $\Sigma ^{H}=-G^{H}W$, instead of the exchange correlation term $V^{\text{xc}}$ in the LDA or the full self-energy $GW$ in the GWA.
We show the $\omega$-dependence of the full self-energy $GW$ and those of the constrained self-energies $G^{L}W$ and $G^{H}W$ in Fig. \ref{KSSigmaSVO} , where the real and imaginary parts of the correlation part at $\Gamma $-point are presented in Fig. \ref{KSSigmaSVO} (a) and (b), respectively.
As one can see in Fig. \ref{KSSigmaSVO}, the $\omega $-dependence of $GW$ mainly originates from $G^{L}W$, and thus the $\omega $-dependence of $G^{H}W$ is very smooth and weak compared to those of $GW$ and $G^{L}W$ and the partial renormalization factor $Z^{H}$ is close to $1$ (see Table \ref{Z_SVO}).
These results are obtained from the $1$st-shot GWA, the first-step of the iterative approximation of the GWA from the DFT/LDA. 
The smoothness of $G^{H}W$ originates partially from the validity of $G^{H}$ as the initial-state of the iterative calculation of the GWA. 
Because of the  smoothness and the small frequency dependence of $G^{H}W$, the truncation up to the first-order expansion of $\Sigma ^{H}$ in frequency in Eq. (\ref{rGexp}) and the neglect of Im$\Sigma ^{H}$ and $Z^{H}$ in the numerator of $G^{Lp}$ in Eq. (\ref{rG2}) are justified.
Actually, $Z^{H}$ in this calculation does not sensitively depend on the choice of the $\omega$ range.   
Figure \ref{tGrWomegaSVO100and110} is the $\omega$-dependence of the transfer integral without double-counting terms to the nearest $[1,0,0]$ and next-nearest $[1,1,0]$ V $t_{2g}$ orbitals.
Similar to the self-energy of the $t_{2g}$ Bloch function in Fig. \ref{KSSigmaSVO}, the $\omega$-dependence of the transfer integral without double-counting terms is very weak.   
In the next paragraph, the effect of the weak $\omega $-dependence is renormalized to the transfer integral through the renormalization factor $Z^{H}$.
As one can see in the list for $Z^H$ in Table \ref{Z_SVO}, the band narrowing effect by the $\omega$ dependence of $\Sigma^{H}$ is small and nearly uniform in $k$-space.

\begin{figure}[tb]
\centering
\includegraphics[clip,width=0.45\textwidth ]{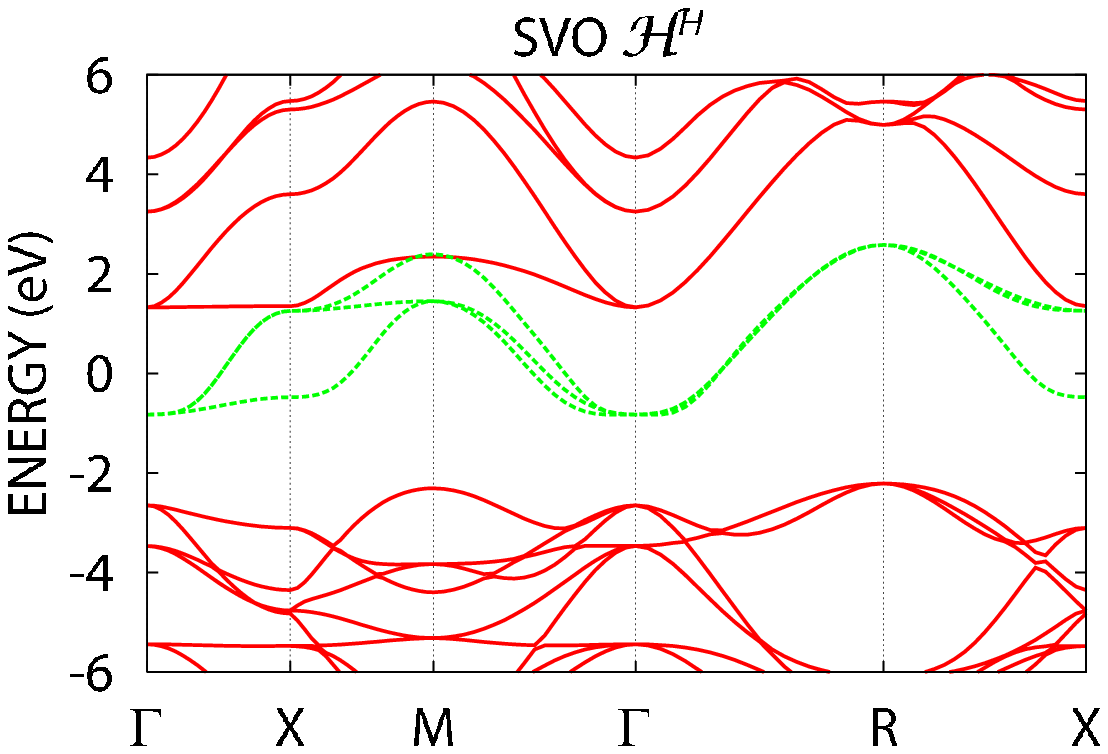}  
\caption{(Color online) Electronic band structure of SVO, where we calculate $\Sigma ^{H}=-G^{H}W$ for $t_{2g}$ Wannier orbitals instead of $V^{\text{xc}}$ according to Eq. (\ref{rG2}), and subtract the average of the self-energy from the low-energy bands. 
The zero energy corresponds to the Fermi level. 
}
\label{bndsSVOGrW}
\end{figure} 
Figure \ref{bndsSVOGrW} shows the band structure after considering the self-energy $G^{H}W$, where we calculate $G^{H}W$ for V $t_{2g}$ bands instead of $V^{\text{xc}}$ and subtract the constant shift of the self-energy from the low-energy bands.
The band structure of the $t_{2g}$ states is calculated according to Eq. (\ref{rG2}), where the effect of the renormalization factor $Z^{H}$ is included.
In Fig. \ref{bndsSVOGrW}, the Fermi level for the low-energy band is determined from the occupation number in the low-energy space.
On the other hand, because the energy is determined not only from the low-energy 
space but also from the high-energy space, the energy is calculated from all 
degree-of-freedom including the high-energy space.
If we calculate the self-energy from the low-energy Green's function $G^L$ in the GWA ($G^{L}W$), then the band structure in the low-energy space  corresponds to that in the full GWA in Fig.~\ref{bndsSVOLDAandGW} (b).    

\begin{table*}[h!tb] 
\caption{Transfer integral and its components for the $t_{2g}$ orbitals of the V sites in the bulk SVO, $t_{mn}(R_x, R_y, R_z)$, 
where $m$ and $n$ denote symmetries of $t_{2g}$ orbitals. Units are given in meV. 
$t^{\text{LDA}}$ is the expectation value of the KS-Hamiltonian for the Wannier function : $t^{\text{LDA}}=\langle \phi^{L}| \mathcal{H}^{\text{LDA}} |\phi^{L} \rangle$ (Eq. (\ref{tLDA})).
$t^{\text{GWA}}$ is the static transfer integral in the GWA : 
$t^{\text{GWA}}=\langle \phi^{L}| \mathcal{H}^{\text{LDA}}+Z(-V^{\text{xc}}+\text{Re}\Sigma) |\phi^{L} \rangle$.
"$V^{\text{xc}}$" is the expectation value of the exchange-correlation potential for the Wannier function : 
"$V^{\text{xc}}$"$=\langle \phi^{L}| V^{\text{xc}} |\phi^{L} \rangle$.
"$(\Sigma _{H}+\Delta\Sigma _{L})(0)$" is the expectation value of the constrained self-energies at $\omega=0$ for the Wannier function : 
"$(\Sigma _{H}+\Delta\Sigma _{L})(0)$"$=\langle \phi^{L}| (\Sigma _{H}+\Delta\Sigma _{L})(\omega=0) |\phi^{L} \rangle$.
$t^{H}$ is the static transfer integral without double-counting : $t^{H}=\langle \phi^{L}| \mathcal{H}^{\text{LDA}}+Z^{H}(-V^{\text{xc}}+\text{Re}\Sigma ^{H}) |\phi^{L} \rangle$  (Eq. (\ref{HGrW})). 
$t^{H\Delta}$ is the static transfer integral including the correction of the $\omega $-dependence of $U$ : $t^{H\Delta}=\langle \phi^{L}|\mathcal{H}^{\text{LDA}}+Z^{H\Delta}(-V^{\text{xc}}+\text{Re}(\Sigma^{H}+\Delta \Sigma ^{L}))|\phi^{L} \rangle$ (Eq. (\ref{HGrWdelta})).
Figure in the bracket in [0,0,0] is the value of the constant shift from the Fermi level of the high-energy space.
 } 
\
\begin{tabular}{c|rrrr|rrrr} 
\hline \hline \\ [-4pt]
SVO & & $t^{\text{LDA}}$ & & & & $t^{\text{GWA}}$ & & \\ [+2pt] 
\hline \\ [-4pt]
$(m, n)$ $\backslash$ $\bm{R}$ 
& \big[0,0,0\big] 
& \big[1,0,0\big] 
& \big[1,1,0\big] 
& \big[1,1,1\big]
& \big[0,0,0\big] 
& \big[1,0,0\big] 
& \big[1,1,0\big]
& \big[1,1,1\big] \\ [+4pt]
\hline \\ [-8pt]
$(xy,xy)$&   684 & $-$271 &  $-$87 & $-$6 &  550(+856) & $-$237 & $-$76 & $-$6 \\
$(xy,yz)$&     0 &      0 &      0 &    4 &    0 &      0 &     0 &    2 \\
$(xy,zx)$&     0 &      0 &      0 &    4 &    0 &      0 &     0 &    2 \\ 
$(yz,yz)$&   684 &  $-$31 &      6 & $-$6 &  550(+856) &  $-$21 &     5 & $-$6 \\
$(yz,zx)$&     0 &      0 &     10 &    4 &    0 &      0 &     9 &    2 \\
$(zx,zx)$&   684 & $-$271 &      6 & $-$6 &  550(+856) & $-$237 &     5 & $-$6 \\ 
\hline \\ [-8pt] 
 & & $V^{\text{xc}}$ &   & &      & $(\Sigma _{H}+\Delta\Sigma _{L})(0)$   & &    \\ [+1pt]
\hline \\ [-8pt] 
$(xy,xy)$& -24355 &    264 &    101 &     7 & $-$21156 & 157 &    71 &     5 \\
$(xy,yz)$&      0 &      0 &      0 & $-$24 &        0 &   0 &     0 & $-$23 \\
$(xy,zx)$&      0 &      0 &      0 & $-$24 &        0 &   0 &     0 & $-$23 \\ 
$(yz,yz)$& -24355 &    134 &  $-$10 &     7 & $-$21156 & 118 &  $-$9 &     5 \\
$(yz,zx)$&      0 &      0 &   $-$4 & $-$24 &        0 &   0 &  $-$2 & $-$23 \\
$(zx,zx)$& -24355 &    264 &  $-$10 &     7 & $-$21156 & 157 &  $-$9 &     5 \\ 
\hline \\ [-8pt] 
 & & $t^{H}$ &   & &      & $t^{H\Delta}$   & &    \\ [+1pt]
\hline \\ [-8pt] 
$(xy,xy)$&  991(+8112) & $-$410 &  $-$64 & $-$2 &  660(+2103) & $-$281 &  $-$86 & $-$6 \\
$(xy,yz)$&           0 &      0 &      0 &   11 &           0 &      0 &      0 &    4 \\
$(xy,zx)$&           0 &      0 &      0 &   11 &           0 &      0 &      0 &    4 \\ 
$(yz,yz)$&  991(+8112) &  $-$66 &   $-$6 & $-$2 &  660(+2103) &  $-$30 &      5 & $-$6 \\
$(yz,zx)$&           0 &      0 &      5 &   11 &           0 &      0 &      9 &    4 \\
$(zx,zx)$&  991(+8112) & $-$410 &   $-$6 & $-$2 &  660(+2103) & $-$281 &      5 & $-$6 \\ 
\hline \hline 
\end{tabular}
\label{t_SVO} 
\end{table*} 

\begin{figure*}[ptb]
\begin{center} 
\includegraphics[width=0.85\textwidth ]{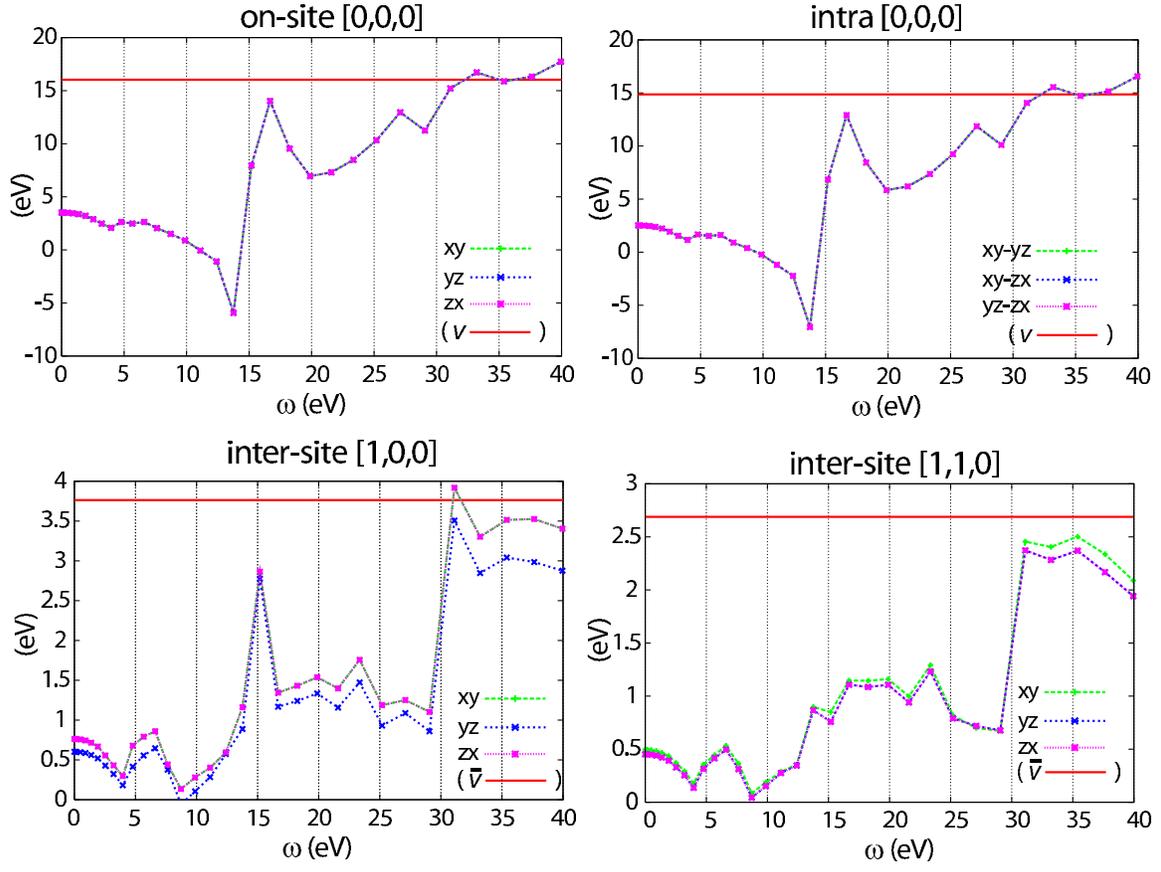} 
\end{center} 
\caption{(Color online) Frequency dependence of partially screened Coulomb interaction $U$ of the $t_{2g}$ orbitals of the V sites. Here, $v$ is the bare Coulomb interaction, and $\bar{v}$ is the average value of the bare Coulomb interaction of the V $t_{2g}$ orbitals.
}
\label{WomegaSVO}
\end{figure*} 

\begin{table*}[ptb] 
\caption{
Effective Coulomb interaction between the two electrons for all the combinations of V $t_{2g}$ orbitals in SVO (in eV).
$v$ and $J_{v}$ represent the bare Coulomb interaction/exchange interactions respectively. $U(0)$ and $J(0)$ represent the static values of the effective Coulomb interaction/exchange interactions (at $\omega=0$). The index 'n' and 'nn' represent the nearest V site [1,0,0] and the next-nearest V site [1,1,0] respectively. 
}
\ 
\label{W_SVO} 
\begin{tabular}{c|ccc|ccc|ccc|ccc|ccc} 
\hline \hline \\ [-8pt]
SVO   &       &  $v$  &       &     & $U(0)$ &    &       & $J_{v}$ &      &     &  $J(0)$ &     \\ [+1pt]
\hline \\ [-8pt]
      &  $xy$ &  $yz$ &  $zx$ & $xy$ & $yz$ & $zx$ &  $xy$ & $yz$ & $zx$ & $xy$ & $yz$ & $zx$ \\ 
\hline \\ [-8pt] 
$xy$  & 16.03 & 14.86 & 14.86 &  3.51 & 2.52 & 2.52 &      & 0.55 & 0.55 &      & 0.47 & 0.47 \\ 
$yz$  & 14.86 & 16.03 & 14.86 &  2.52 & 3.51 & 2.52 & 0.55 &      & 0.55 & 0.47 &      & 0.47 \\
$zx$  & 14.86 & 14.86 & 16.03 &  2.52 & 2.52 & 3.51 & 0.55 & 0.55 &      & 0.47 & 0.47 &      \\
\hline \hline \\ [-8pt]  
      &       &  $v_{\text{n}}$ &    &     & $V_{\text{n}}(0)$ &    &       & $v_{\text{nn}}$  &      &     &  $V_{\text{nn}}(0)$ &     \\ [+1pt]
\hline \\ [-8pt] 
      &  $xy$ &  $yz$ &  $zx$ & $xy$ & $yz$ & $zx$ &  $xy$ & $yz$ & $zx$ & $xy$ & $yz$ & $zx$ \\ 
\hline \\ [-8pt] 
$xy$  &  3.90 &  3.67 &  3.88 & 0.76 & 0.67 & 0.74 &  2.77 & 2.71 & 2.71 & 0.50 & 0.47 & 0.47 \\ 
$yz$  &  3.67 &  3.48 &  3.67 & 0.67 & 0.60 & 0.67 &  2.71 & 2.65 & 2.66 & 0.47 & 0.45 & 0.46 \\
$zx$  &  3.88 &  3.67 &  3.90 & 0.74 & 0.67 & 0.76 &  2.71 & 2.66 & 2.65 & 0.47 & 0.46 & 0.45 \\
\hline
\hline 
\end{tabular} 
\end{table*}

We summarize the values of the transfer integral in Table \ref{t_SVO}.
The transfer $t^{H}$ in the left column of Table \ref{t_SVO} corresponds to Fig. \ref{bndsSVOGrW}, where the $\omega$-dependence of $\Sigma ^{H}$ is renormalized to $Z^{H}$.
The renormalized values obtained further from the $\omega$-dependence of the partial screened Coulomb interaction $W^{H}(\omega)$, namely, $\Delta \Sigma^{L}$ and $t^{H\Delta}$, are discussed in the next paragraph.
We note that the parameter of the $t_{2g}$ orbital has the same symmetry with the V $t_{2g}$ orbitals in the real SVO. 
In the LDA, the largest value of the nearest hopping is between the same symmetries $t^{\text{LDA}}_{xy,xy}(1,0,0)$ and is about 270 meV, whereas that of the next nearest hopping $t^{\text{LDA}}_{xy,xy}(1,1,0)$ is about $90$ meV.   
Compared to the results of the LDA and the full GWA in Fig. \ref{bndsSVOLDAandGW}, the band width of the V $t_{2g}$ in Fig. \ref{bndsSVOGrW} is much larger because of the absence of the self-energy originating from the low-energy space $G^{L}W$ (see Table \ref{width_SVO}).
Correspondingly, the transfer integral for the nearest-neighbor pair becomes about $50$\% larger than that of the LDA. 
In other words, the absolute value of the ``effective" nearest transfer integral is about $130$ meV lower than that of the ``bare" nearest transfer integral given by $t^{\text{LDA}}-V^{\text{xc}}$ because of the screening effect of the high-energy degrees of freedom.

\begin{figure}[ptb]
\centering 
\includegraphics[width=0.45\textwidth ]{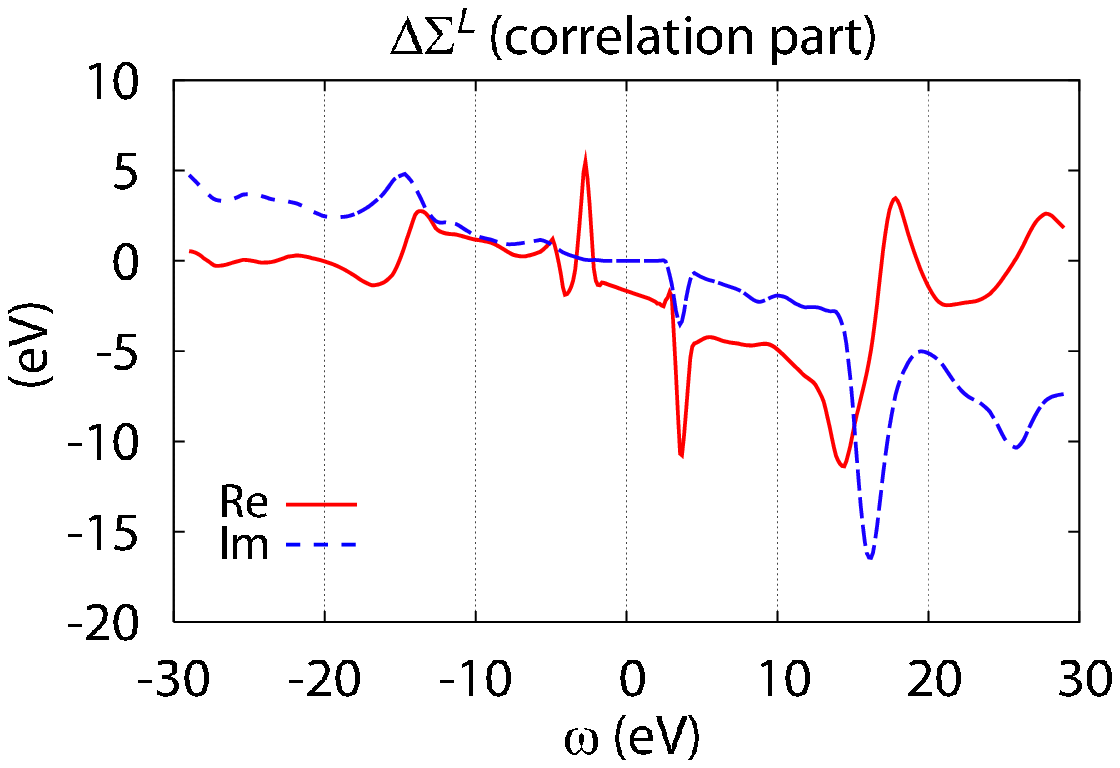}  
\caption{(Color online) Correlation part of $\Delta \Sigma ^{L}$ of the $t_{2g}$ Bloch functions of SVO at $\Gamma=(0,0,0)$ in the Kohn-Sham gauge. 
}
\label{KSdeltaSigmaSVO}
\end{figure} 
\begin{figure}[ptb]
\centering
\includegraphics[width=0.45\textwidth ]{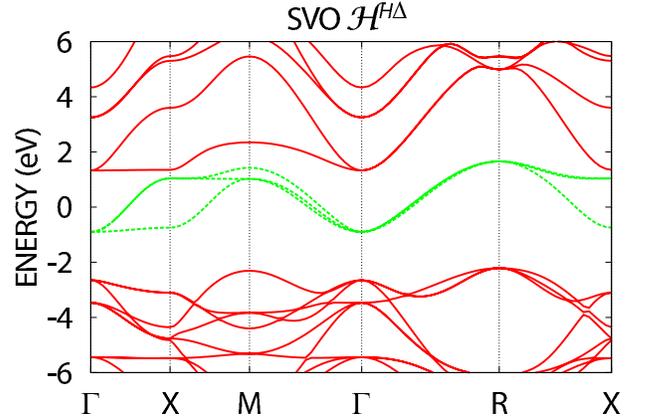} 
\caption{(Color online) Electronic band structure of SVO, where we calculate $\Sigma ^{H}+\Delta \Sigma ^{L}$ for $t_{2g}$ Wannier orbitals instead of $V^{\text{xc}}$ according to Eq. (\ref{rdelG2}), and subtract the average of the self-energy from the low-energy bands. 
The zero energy corresponds to the Fermi level. 
}
\label{bndsSVOGrWdelta}
\end{figure} 

We now consider the self-energy effect arising from the $\omega$-dependence of the partially screened Coulomb interaction.
Figure \ref{WomegaSVO} is the $\omega$-dependence of the partially screened Coulomb interaction between the V $t_{2g}$ orbitals.
As is the case with the constrained self-energy $\Sigma^{H}$, the $\omega$-dependence of the partially screened Coulomb interaction $W^{p}$ is weak for small $\omega (<10$ eV). 
We summarize the static limit of the partially screened Coulomb interaction in Table \ref{W_SVO}.
We renormalize the $\omega$-dependence of the screened Coulomb interaction neglected in Table \ref{W_SVO} into the one-body part by following Eq. (\ref{dsig}).
Figure \ref{KSdeltaSigmaSVO} shows the $\omega$-dependence of the correlation term of $\Delta \Sigma ^{L}=-G^{L}(W-W^{U})$. 
Because $\Delta \Sigma ^{L}$ originates from the low-energy degrees of freedom, the overall slope in the $\omega$-dependence of $\Delta \Sigma ^{L}$ is basically similar to $G^{L}W$ illustrated in Fig.\ref{KSSigmaSVO}(a).  
Small dip originating from $L$-$L$ polarization $G^{L}G^{L}$ is observed near the energy eigenvalue of the low-energy band.

Figure \ref{bndsSVOGrWdelta} illustrates the band structure where the $\omega$ dependent part of the partially screened Coulomb interaction is renormalized into the one-body part by following Eq. (\ref{rdelG2}).
This band not only includes the effect of $Z^{\Delta}= (1-\frac{\partial \Delta \Sigma ^{L}}{\partial \omega})^{-1}$ but also includes the effect of $Z^{H}$.     
In Table \ref{t_SVO}, the nearest transfer integral renormalized by the $\omega$-dependence of $U$, $t^{H\Delta}$, is substantially reduced from $t^{H}$.
This reduction of the transfer integral is mainly caused by the renormalization factor $Z^{H\Delta}$.

It is remarkable that the resultant band structure of the effective model is nearly the same as that of the LDA. 
Table \ref{width_SVO} shows that the final band width is 2.56 eV in comparison to the LDA result of 2.58 eV. 
The two corrections originated from different effects, namely the increase in the band width arising from $\Sigma^{H}$ and the reduction arising from $\Delta \Sigma ^{L}$ are roughly compensated and results in $t^{H\Delta}$ similar to the LDA estimate.


\subsection{FeSe and FeTe}

\begin{figure*}[ptb]
\begin{center} 
\includegraphics[width=0.9\textwidth ]{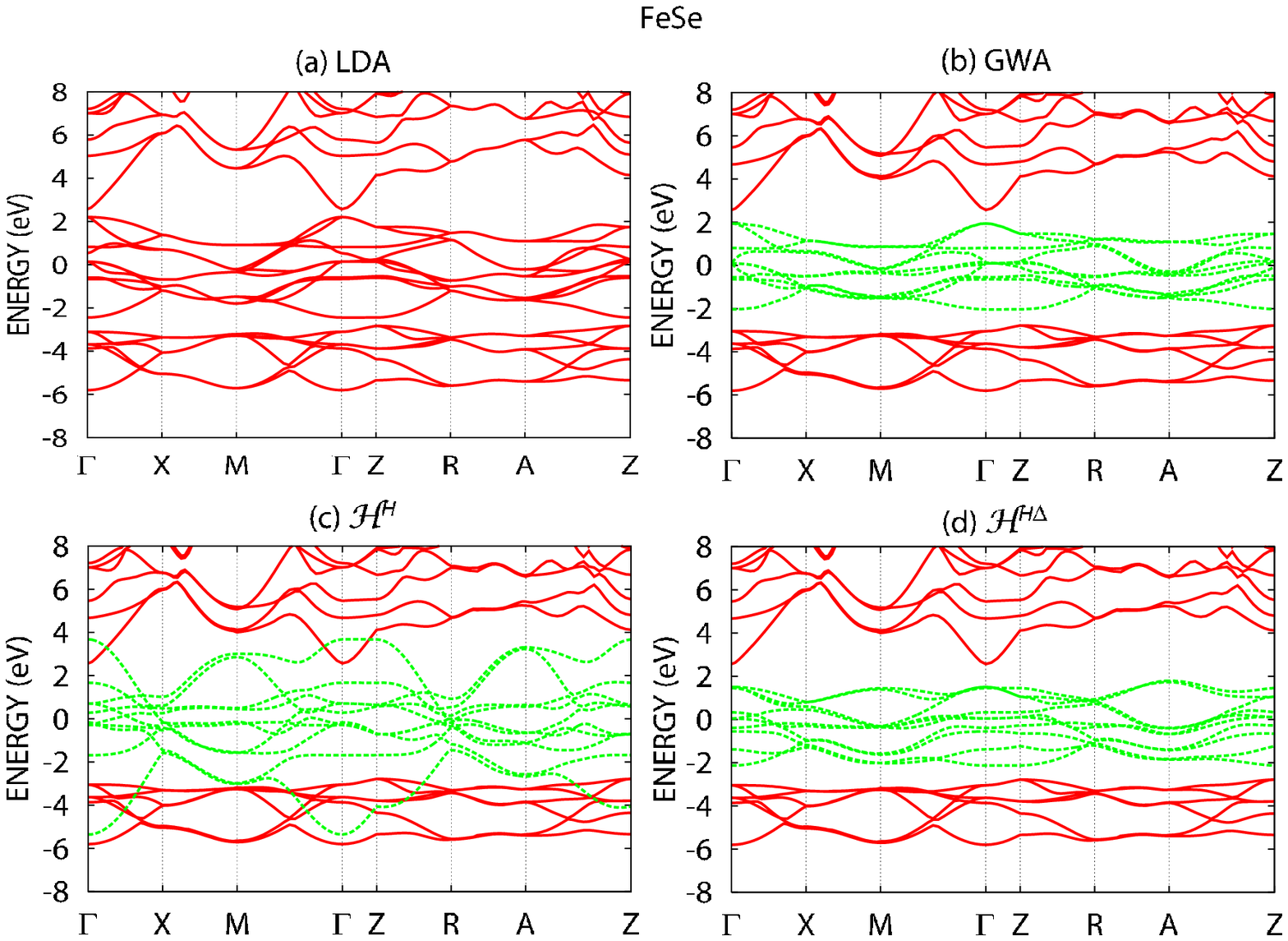} 
\end{center} 
\caption{(Color online) Electronic band structures of FeSe obtained by (a) the LDA and (b) the GWA and calculated according to (c) Eq. (\ref{rG2}) (double-counting-less of the low-energy space) and (d) Eq. (\ref{rdelG2}) (renormalizing the $\omega$-dependence of $W^{p}$). The zero energy corresponds to the Fermi level. }
\label{bndsFeSe}
\end{figure*} 
\begin{figure}[ptb]
\centering 
\includegraphics[clip,width=0.38\textwidth ]{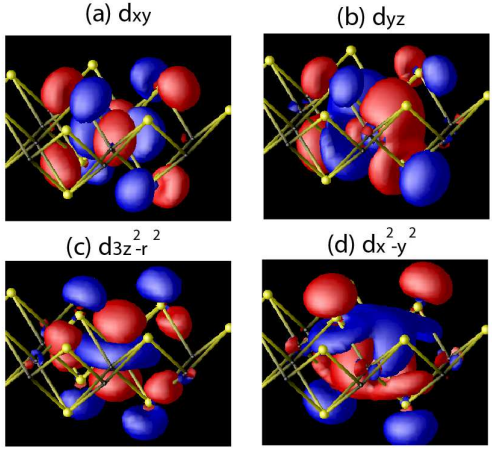} 
\caption{(Color online) Isosurface of the maximally localized Wannier functions $\pm 0.02$ a.u. for the Fe $3d$ orbitals in FeSe. }
\label{WannierFeSe}
\end{figure} 
\begin{figure*}[htb]
\begin{center} 
\includegraphics[width=0.9\textwidth ]{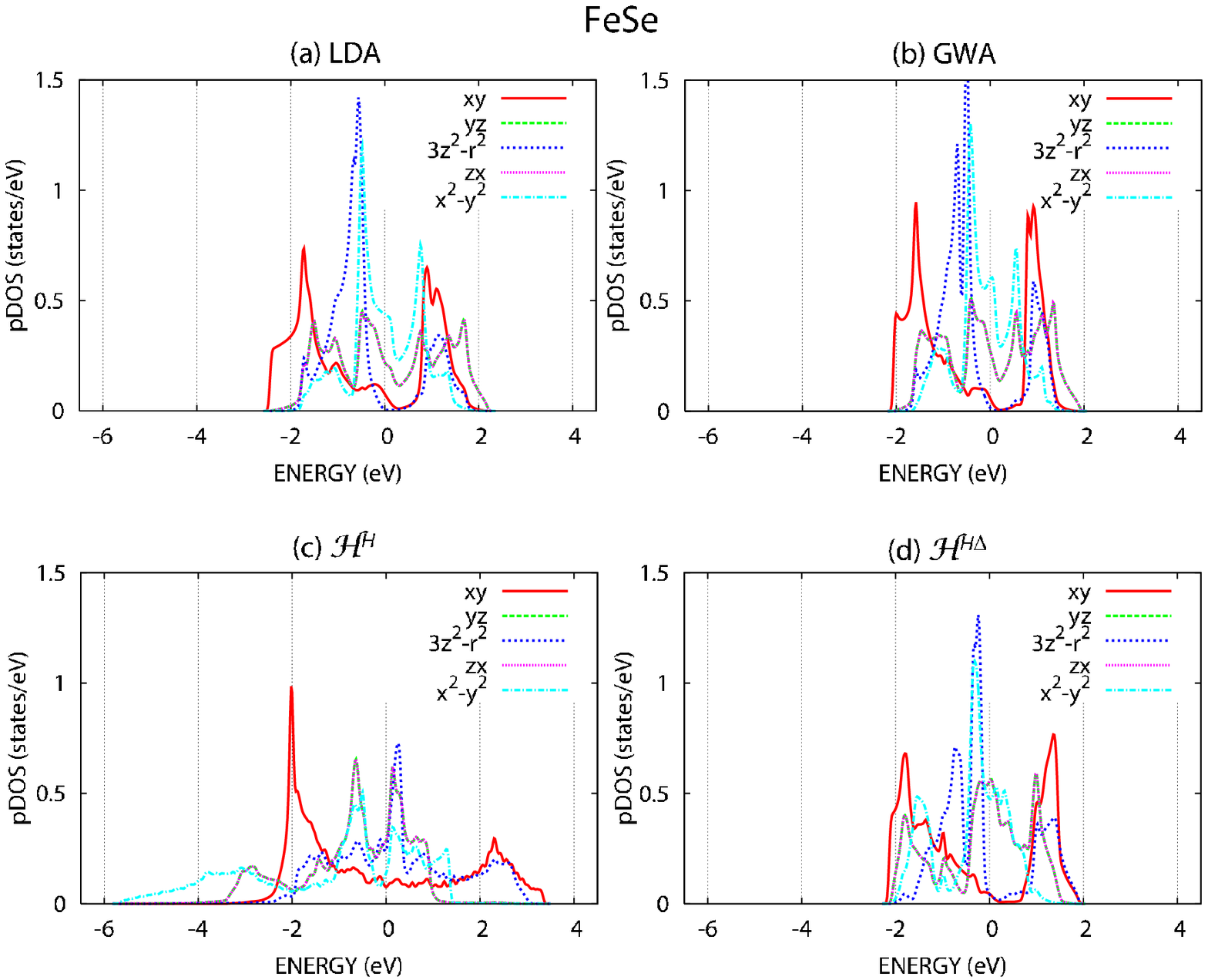} 
\end{center} 
\caption{(Color online) Partial densities of FeSe resolved by the Wannier function of the Fe $3d$ orbital: (a) the LDA and (b) the GWA and (c) according to Eq. (\ref{rG2}) (double-counting-less of the low-energy space) and (d) according to Eq. (\ref{rdelG2}) (renormalizing the $\omega$-dependence of $W^{p}$). The zero energy corresponds to the Fermi level. }
\label{DOSFeSe}
\end{figure*} 

FeSe$_{x}$Te$_{1-x}$ is the simplest iron-based superconductor\cite{hosono}. It shows superconductivity with the transition temperature $T_{c}\sim 10$ K at ambient pressure\cite{hsu,li} and $T_{c}\sim 37$ K under pressure ($7$ GPa)\cite{margadonna}. 
FeTe indicates an antiferromagnetic ordered moment $\sim 2.0$-$2.25 \mu_{\text{B}}$ at a Bragg point ($\pi /2, \pi /2$) in the extended Brillouin zone\cite{li, bao}, whereas magnetism is not observed in FeSe. 
In this subsection, we derive the effective model of the two compounds and discuss how orbital-dependent effective parameters affect low-energy properties. 

Figure \ref{bndsFeSe} (a) shows the band structure of FeSe in the DFT/LDA. 
As is the general case in iron-based superconductors \cite{miyake2}, 
ten states having strong Fe 3$d$ character form a band near the Fermi level. The band is about $4.5$ eV in 
width, and is occupied by twelve electrons per unit cell. 
Small electron pockets are found around the M point and hole pockets are around the $\Gamma$ point. 
Below the $d$ bands, six states exist between $-6$ and $-3$ eV. They consist of mainly Se 4$p$ orbitals.

We regard the ten states having Fe $3d$ character as the low-energy states, and construct the effective model.
Figure \ref{WannierFeSe} shows the isosurface of the maximally localized Wannier functions associated with 
the ten states. They are spatially extended 
because of hybridization between the Fe $3d$ atomic orbital and adjacent Se $4p$ atomic orbitals. 
Since the strength of hybridization depends on the orbital, spread of the Wannier functions 
is orbital-dependent (Table \ref{spread_iron}). 
This is in sharp contrast with the $t_{2g}$-Wannier orbitals of SVO in the previous section. 
The $yz$/$zx$ and $x^{2}-y^{2}$ orbitals are delocalized compared to the $xy$ and $3z^{2}-r^{2}$ orbitals. 
This trend is observed not only in FeSe but also in FeTe and other iron-based superconductors \cite{miyake2}. 
(The $xy$ axes in our convention are along the unit vectors of the cell containing two Fe atoms\cite{miyake2}. 
We abbreviate the $3d$ orbitals such as $d_{xy}$ as $xy$, unless confusions occur.)

Orbital dependence is also observed in the partial density of states (pDOS). 
Figure \ref{DOSFeSe} presents the pDOS of FeSe resolved by the Wannier functions. 
The $x^2-y^2$ orbital has large density of states at the Fermi level in the LDA. 
A dip (pseudogap) is seen at about $0.3$ eV above the Fermi level. 
Table \ref{Occ_FeSe} shows the occupation number, where the $3z^{2}-r^{2}$ orbital has the largest value (nearly $3/4$-filling), while the $x^{2}-y^{2}$ and $yz$/$zx$ orbitals are about half-filling.
We will discuss the results other than LDA later.

Figures \ref{bndsFeSe} (b) is the band structure of FeSe in the GWA, where we add the self-energy to the low-energy states 
after removing the LDA exchange-correlation potential. 
The band dispersion is similar to the LDA result, though 
the bandwidth becomes about 10\% smaller by the self-energy effect. 
This trend is more clearly seen in the pDOS.  In Fig. \ref{DOSFeSe} (b), we can confirm that the pDOS is 
similar to that of LDA except for overall band narrowing. The occupation number of each orbital 
is nearly the same as the LDA value. 

\begin{figure*}[tb]
\begin{center} 
\includegraphics[width=0.8\textwidth ]{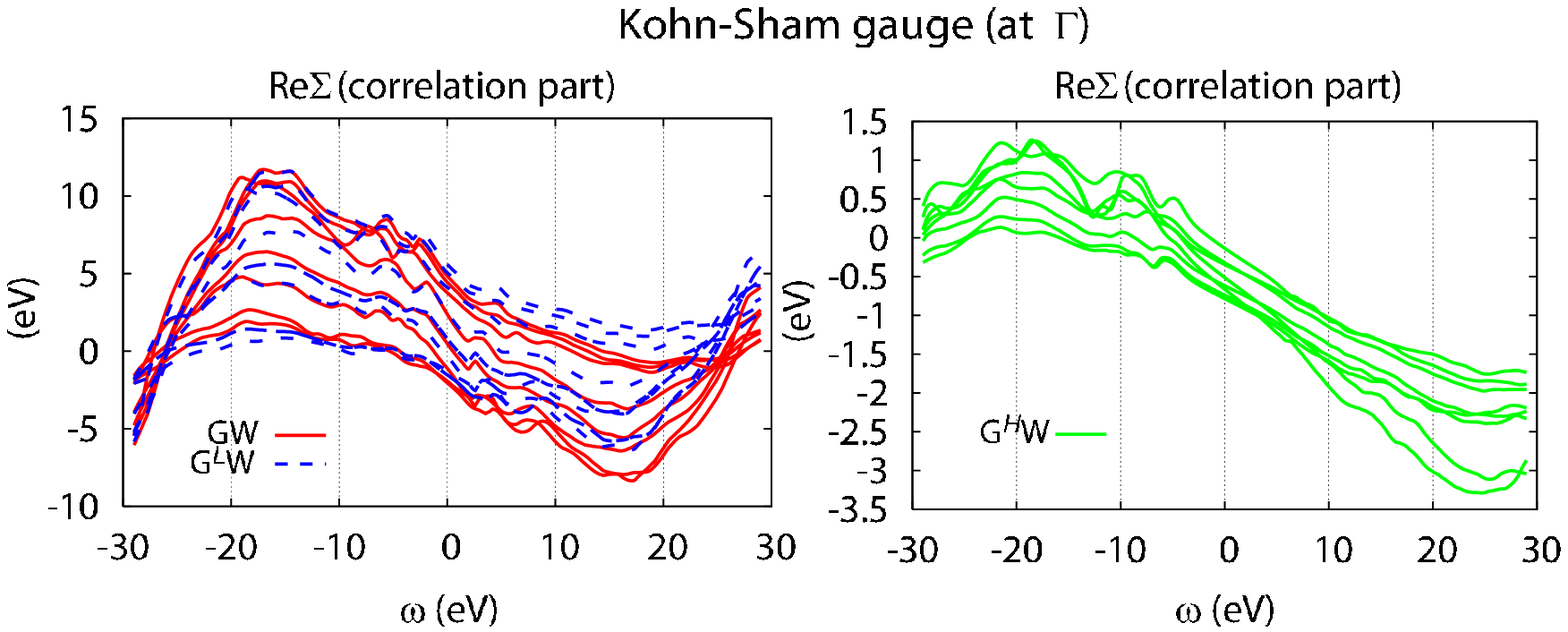} 
\end{center} 
\caption{Real parts of correlation self-energy of the $3d$ Bloch functions at $\Gamma=(0,0,0)$ in the Kohn-Sham gauge.}
\label{KSSigmaReFeSe}
\end{figure*} 
\begin{figure*}[tb]
\begin{center} 
\includegraphics[width=0.8\textwidth ]{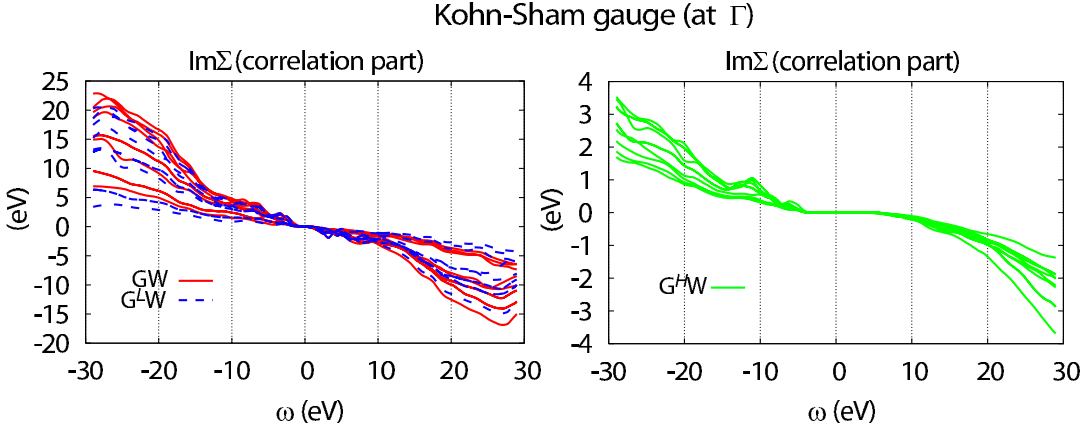} 
\end{center} 
\caption{Imaginary parts of correlation self-energy of the $3d$ Bloch functions at $\Gamma=(0,0,0)$ in the Kohn-Sham gauge.}
\label{KSSigmaImFeSe}
\end{figure*} 

\begin{table}[tb] 
\caption{Spread of the Wannier functions (in \AA $^{2}$) defined by quadratic extent.
} 
\
\begin{tabular}{c|ccccc}
\hline \hline \\ [-8pt]  
Spread               & $xy$ & $yz$ & $3z^{2}-r^{2}$ & $zx$ & $x^{2}-y^{2}$   \\ [+1pt]
\hline \\ [-8pt] 
FeSe                 & 1.66 & 1.94 & 1.58 & 1.94 & 2.36    \\ 
FeTe                 & 2.35 & 2.76 & 2.01 & 2.76 & 2.43    \\ 
\hline \hline 

\end{tabular}
\label{spread_iron} 
\end{table}

\begin{table}[tb] 
\caption{Occupation number of the Wannier functions of FeSe, where the sum of the occupancy is $6$.
} 
\
\begin{tabular}{c|ccccc}
\hline \hline \\ [-8pt]  
Occ. Num.               & $xy$ & $yz$ & $3z^{2}-r^{2}$ & $zx$ & $x^{2}-y^{2}$   \\ [+1pt]
\hline \\ [-8pt] 
LDA                     & 1.23 & 1.05 & 1.57 & 1.05 & 1.11    \\ 
GWA                     & 1.22 & 1.03 & 1.53 & 1.03 & 1.19    \\ 
$\mathcal{H}^{H}$       & 1.18 & 1.30 & 0.86 & 1.30 & 1.36    \\ 
$\mathcal{H}^{H\Delta}$ & 1.19 & 1.01 & 1.43 & 1.01 & 1.35    \\ 
\hline \hline 

\end{tabular}
\label{Occ_FeSe} 
\end{table}

Now we turn to the results of effective models. 
Figure \ref{bndsFeSe} (c) shows the band structure for Eq. (\ref{HGrW}), where the self-energy  correction 
from high-energy states is included.
As is observed in SVO, the band is substantially wider than that of LDA. 
In contrast with SVO, however, the band dispersion is qualitatively different from the LDA band structure.
Both the small electron pockets around the M point and hole pockets around the $\Gamma$ point 
disappear. We note here that this band structure does not include the self-energy effect from the low-energy space, 
hence it should not be compared to experimental measurements.
The partial density of states and occupation number corresponding to Fig. \ref{bndsFeSe} (c) 
are shown in Fig. \ref{DOSFeSe} (c) and Table \ref{Occ_FeSe}, respectively.
The psudogap observed in both LDA and GWA disappears. 
The $x^2-y^2$ orbital has a weight down to $\sim$ $-6$ eV below the Fermi level, while the $xy$ component 
extends up to 3 eV, which results in the band widening.  
The occupation number of the $yz$, $zx$ and $x^2-y^2$ increases, whereas that of the $3z^2-r^2$ decreases 
substantially to 0.86 from the LDA value of 1.57. 
Consequently, the order of the occupation numbers is completely different from the LDA and GWA.

From now on, we focus on the effective model parameters obtained from ${\cal H}^{H\Delta}$ in Eq. (\ref{HGrWdelta}), because they are the best parameters which should be used in low-energy solvers. In "$\mathcal{H}^{H\Delta}$", the $\omega$ dependence of the screened Coulomb interaction is renormalized to the one-body part.
The static value of the on-site screened Coulomb ($U$) and exchange ($J$) interactions in FeSe are presented in 
Table \ref{W_FeSe}. We can see that the $U$ strongly depends on the orbital. 
Inclusion of its $\omega$-dependence using Eq. (\ref{HGrWdelta}) 
yields the band structure shown in Fig. \ref{bndsFeSe} (d). 
The band becomes narrower by the dynamical effect of $U$. 
The bandwidth is about 4 eV, which is smaller than the LDA bandwidth and close to the GW one. 
The pDOS is shown in Fig. \ref{DOSFeSe} (d). 
The $yz/zx$ and $3z^{2}-r^{2}$ components have a peak near the Fermi energy. 
This explains why the numbers are quite different between LDA, GWA, $\mathcal{H}^{H}$ and $\mathcal{H}^{H\Delta}$.

\begin{figure}[tb]
\begin{center} 
\includegraphics[width=0.4\textwidth ]{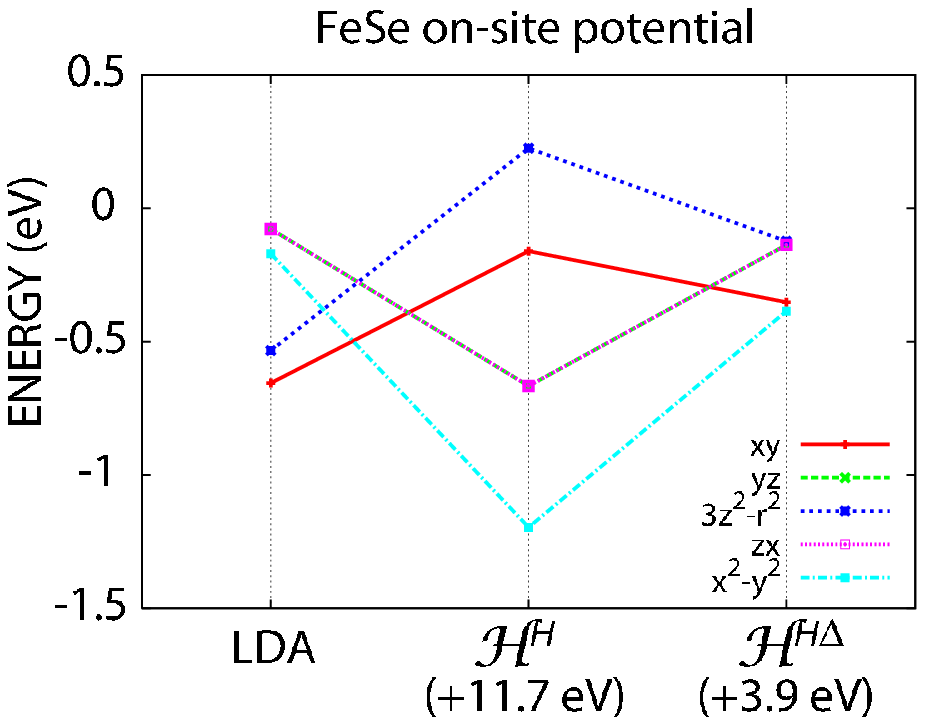} 
\end{center} 
\caption{On-site potential of the Wannier function of FeSe.
Figure in the bracket is the value of the constant shift from the Fermi level of the high-energy space.}
\label{tonFeSe}
\end{figure} 
\begin{figure}[tb]
\begin{center} 
\includegraphics[width=0.45\textwidth ]{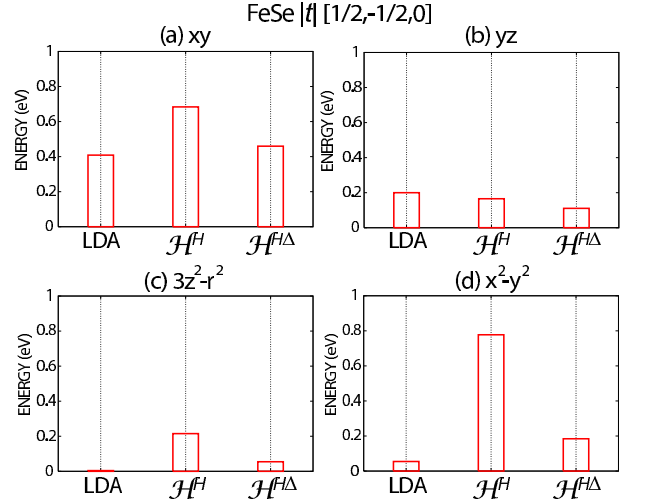} 
\end{center} 
\caption{Nearest-neighbor transfer integral of the Wannier function of FeSe.}
\label{tnnFeSe}
\end{figure} 

\begin{table*}[ptb] 
\caption{Transfer integral and its components for the $3d$ orbitals of the Fe sites in the FeSe, $t_{mn}(R_x, R_y, R_z)$, 
where $m$ and $n$ denote symmetries of $3d$ orbitals. Units are given in meV.  
$t^{\text{LDA}}$ is the expectation value of the KS-Hamiltonian for the Wannier function : $^{\text{LDA}}=\langle \phi^{L}| \mathcal{H}^{\text{LDA}} |\phi^{L} \rangle$ (Eq. (\ref{tLDA})).   
$t^{H}$ is the static transfer integral without double-counting : $t^{H}=\langle \phi^{L}| \mathcal{H}^{\text{LDA}}+Z^{H}(-V^{\text{xc}}+\text{Re}\Sigma ^{H}) |\phi^{L} \rangle$  (Eq. (\ref{HGrW})). 
$t^{H\Delta}$ is the static transfer integral including the correction of the $\omega $-dependence of $U$ : $t^{H\Delta}=\langle \phi^{L}|\mathcal{H}^{\text{LDA}}+Z^{H\Delta}(-V^{\text{xc}}+\text{Re}(\Sigma^{H}+\Delta \Sigma ^{L}))|\phi^{L} \rangle$ (Eq. (\ref{HGrWdelta})).
Figure in the bracket in [0,0,0] is the value of the constant shift from the Fermi level of the high-energy space.
} 
\
\begin{tabular}{c|rrrrrrr|rrr} 
\hline \hline \\ [-4pt]
  $t^{\text{LDA}}$  \\ [+2pt] 
\hline \\ [-4pt]
$(m, n)$ $\backslash$ $\bm{R}$ 
& \big[$0,0,0$\big] 
& \big[$\frac{1}{2},-\frac{1}{2},0$\big] 
& \big[$1,0,0$\big] 
& \big[$1,-1,0$\big] 
& \big[$\frac{3}{2},-\frac{1}{2},0$\big]
& \big[$0,0,\frac{c}{a}$\big] 
& \big[$\frac{1}{2},-\frac{1}{2},\frac{c}{a}$\big]
& $\sigma_{y}$
& $I$
& $\sigma^{L}$ \\ [+4pt]
\hline \\ [-8pt]             
$(xy,xy)$                    & $-$655 & $-$409 &  $-$59 & $-$15 &     6 & $-$25 &    8 & $+$ & $+$ &      $+$ \\ 
$(xy,yz)$                    &      0 &    275 &    134 &  $-$9 &     0 &     0 & $-$7 & $+$ & $-$ & $-$(1,4) \\
$(xy,3z^{2}-r^{2})$          &      0 & $-$347 &      0 &    22 &     0 &     0 &   12 & $-$ & $+$ &      $+$ \\ 
$(xy,zx)$                    &      0 &    275 &      0 &  $-$9 &    37 &     0 & $-$2 & $-$ & $-$ & $-$(1,2) \\
$(xy,x^{2}-y^{2})$           &      0 &      0 &      0 &     0 &     0 &     0 & $-$2 & $-$ & $+$ &      $-$ \\ 
$(yz,yz)$                    &  $-$77 &    200 &    122 & $-$14 & $-$15 &    10 &   31 & $+$ & $+$ &    (4,4) \\ 
$(yz,3z^{2}-r^{2})$          &      0 & $-$119 &      0 &     9 &     4 &     0 &    9 & $-$ & $-$ & $-$(4,3) \\ 
$(yz,zx)$                    &      0 &    130 &      0 & $-$25 &     0 &     0 &   10 & $-$ & $+$ &    (4,2) \\ 
$(yz,x^{2}-y^{2})$           &      0 &    219 &      0 &  $-$2 &  $-$4 &     0 &   19 & $-$ & $-$ &    (4,5) \\ 
$(3z^{2}-r^{2},3z^{2}-r^{2})$& $-$534 &      4 &  $-$34 & $-$16 & $-$13 & $-$23 & $-$6 & $+$ & $+$ &      $+$ \\ 
$(3z^{2}-r^{2},zx)$          &      0 &    119 &    179 &  $-$9 &     0 &     0 & $-$8 & $+$ & $-$ & $-$(3,2) \\ 
$(3z^{2}-r^{2},x^{2}-y^{2})$ &      0 &      0 & $-$105 &     0 &     0 &  $-$7 & $-$3 & $+$ & $+$ &      $-$ \\ 
$(zx,zx)$                    &  $-$77 &    200 &    400 & $-$14 &    28 &    10 &    1 & $+$ & $+$ &    (2,2) \\ 
$(zx,x^{2}-y^{2})$           &      0 & $-$219 &     90 &     2 &     0 &     0 &    6 & $+$ & $-$ &    (2,5) \\ 
$(x^{2}-y^{2},x^{2}-y^{2})$  & $-$171 &  $-$55 &     68 &  $-$9 &    36 & $-$30 &    8 & $+$ & $+$ &      $+$ \\ 
\hline \hline 
  $t^{H}$    \\ [+2pt] 
\hline \\ [-4pt]
$(m, n)$ $\backslash$ $\bm{R}$ 
& \big[$0,0,0$\big] 
& \big[$\frac{1}{2},-\frac{1}{2},0$\big] 
& \big[$1,0,0$\big] 
& \big[$1,-1,0$\big] 
& \big[$\frac{3}{2},-\frac{1}{2},0$\big]
& \big[$0,0,\frac{c}{a}$\big] 
& \big[$\frac{1}{2},-\frac{1}{2},\frac{c}{a}$\big]
& $\sigma_{y}$
& $I$
& $\sigma^{L}$ \\ [+4pt]
\hline \\ [-8pt]             
$(xy,xy)$                    &  $-$160(+11716) & $-$684 &     93 &   113 &     19 & $-$60 &     7 & $+$ & $+$ &      $+$ \\ 
$(xy,yz)$                    &       0 &     45 &     97 & $-$37 &      0 &     0 &     2 & $+$ & $-$ & $-$(1,4) \\
$(xy,3z^{2}-r^{2})$          &       0 & $-$517 &      0 &    42 &      0 &     0 &     3 & $-$ & $+$ &      $+$ \\ 
$(xy,zx)$                    &       0 &     45 &      0 & $-$37 & $-$120 &     0 & $-$10 & $-$ & $-$ & $-$(1,2) \\
$(xy,x^{2}-y^{2})$           &       0 &      0 &      0 &     0 &      0 &     0 & $-$23 & $-$ & $+$ &      $-$ \\ 
$(yz,yz)$                    &  $-$666(+11716) & $-$165 &    171 & $-$48 &  $-$54 &     8 &    37 & $+$ & $+$ &    (4,4) \\ 
$(yz,3z^{2}-r^{2})$          &       0 &  $-$25 &      0 &    31 &      8 &     0 &  $-$3 & $-$ & $-$ & $-$(4,3) \\ 
$(yz,zx)$                    &       0 &    410 &      0 & $-$76 &      0 &     0 &    11 & $-$ & $+$ &    (4,2) \\ 
$(yz,x^{2}-y^{2})$           &       0 &    152 &      0 & $-$18 &     40 &     0 &     1 & $-$ & $-$ &    (4,5) \\ 
$(3z^{2}-r^{2},3z^{2}-r^{2})$&     226(+11716) & $-$215 & $-$214 &    59 &     96 &    51 &    35 & $+$ & $+$ &      $+$ \\ 
$(3z^{2}-r^{2},zx)$          &       0 &     25 &    156 & $-$31 &      0 &     0 &    21 & $+$ & $-$ & $-$(3,2) \\ 
$(3z^{2}-r^{2},x^{2}-y^{2})$ &       0 &      0 &    146 &     0 &  $-$78 & $-$62 &    42 & $+$ & $+$ &      $-$ \\ 
$(zx,zx)$                    &  $-$666(+11716) & $-$165 &    119 & $-$48 &     78 &     8 &  $-$3 & $+$ & $+$ &    (2,2) \\ 
$(zx,x^{2}-y^{2})$           &       0 & $-$152 &    275 &    18 &      0 &     0 & $-$13 & $+$ & $-$ &    (2,5) \\ 
$(x^{2}-y^{2},x^{2}-y^{2})$  & $-$1197(+11716) &    778 & $-$110 & $-$20 &  $-$20 & $-$62 &    47 & $+$ & $+$ &      $+$ \\ 
\hline \hline 
  $t^{H\Delta}$    \\ [+2pt] 
\hline \\ [-4pt]
$(m, n)$ $\backslash$ $\bm{R}$ 
& \big[$0,0,0$\big] 
& \big[$\frac{1}{2},-\frac{1}{2},0$\big] 
& \big[$1,0,0$\big] 
& \big[$1,-1,0$\big] 
& \big[$\frac{3}{2},-\frac{1}{2},0$\big]
& \big[$0,0,\frac{c}{a}$\big] 
& \big[$\frac{1}{2},-\frac{1}{2},\frac{c}{a}$\big]
& $\sigma_{y}$
& $I$
& $\sigma^{L}$ \\ [+4pt] 
\hline \\ [-8pt]             
$(xy,xy)$                    & $-$352(+3916) & $-$460 &  $-$16 &    19 &     8 & $-$41 &     9 & $+$ & $+$ &      $+$ \\ 
$(xy,yz)$                    &      0 &    204 &    103 & $-$25 &     0 &     0 &  $-$3 & $+$ & $-$ & $-$(1,4) \\
$(xy,3z^{2}-r^{2})$          &      0 & $-$383 &      0 &    20 &     0 &     0 &    12 & $-$ & $+$ &      $+$ \\ 
$(xy,zx)$                    &      0 &    204 &      0 & $-$25 & $-$10 &     0 &  $-$7 & $-$ & $-$ & $-$(1,2) \\
$(xy,x^{2}-y^{2})$           &      0 &      0 &      0 &     0 &     0 &     0 & $-$10 & $-$ & $+$ &      $-$ \\ 
$(yz,yz)$                    & $-$137(+3916) &    110 &    146 & $-$33 & $-$33 &    10 &    39 & $+$ & $+$ &    (4,4) \\ 
$(yz,3z^{2}-r^{2})$          &      0 &  $-$87 &      0 &    22 &     3 &     0 &     4 & $-$ & $-$ & $-$(4,3) \\ 
$(yz,zx)$                    &      0 &    190 &      0 & $-$33 &     0 &     0 &    12 & $-$ & $+$ &    (4,2) \\ 
$(yz,x^{2}-y^{2})$           &      0 &    194 &      0 &  $-$2 &    21 &     0 &    10 & $-$ & $-$ &    (4,5) \\ 
$(3z^{2}-r^{2},3z^{2}-r^{2})$& $-$123(+3916) &  $-$55 & $-$111 &    21 &    16 &  $-$5 &     5 & $+$ & $+$ &      $+$ \\ 
$(3z^{2}-r^{2},zx)$          &      0 &     87 &    162 & $-$22 &     0 &     0 &     1 & $+$ & $-$ & $-$(3,2) \\ 
$(3z^{2}-r^{2},x^{2}-y^{2})$ &      0 &      0 &  $-$29 &     0 & $-$35 & $-$24 &     9 & $+$ & $+$ &      $-$ \\ 
$(zx,zx)$                    & $-$137(+3916) &    110 &    312 & $-$33 &    74 &    10 &     3 & $+$ & $+$ &    (2,2) \\ 
$(zx,x^{2}-y^{2})$           &      0 & $-$194 &    131 &     2 &     0 &     0 &  $-$2 & $+$ & $-$ &    (2,5) \\ 
$(x^{2}-y^{2},x^{2}-y^{2})$  & $-$385(+3916) &    184 &     13 & $-$32 &    18 & $-$38 &    18 & $+$ & $+$ &      $+$ \\ 
\hline \hline 
\end{tabular}
\label{t_FeSe} 
\end{table*}

\begin{table*}[htb] 
\caption{
On-site effective Coulomb interaction between the two electrons for all the combinations of Fe $3d$ orbitals in FeSe (in eV).
$U(0)$ and $J(0)$ represent the static values of the on-site effective Coulomb interaction/exchange interactions (at $\omega=0$). 
}
\ 
\label{W_FeSe} 
\begin{tabular}{c|ccccc|ccccc} 
\hline \hline \\ [-8pt]
FeSe           &      &      &     $v$     &      &               &      &      &      $U(0)$    &      &          \\ [+1pt]
\hline \\ [-8pt]
               & $xy$ & $yz$ & $3z^{2}-r^{2}$ & $zx$ & $x^{2}-y^{2}$ & $xy$ & $yz$ & $3z^{2}-r^{2}$ & $zx$ & $x^{2}-y^{2}$ \\ 
\hline \\ [-8pt] 
$xy$           & 18.69 & 16.57 & 17.27 & 16.57 & 16.51  & 4.60 & 3.28 & 3.29 & 3.28 & 3.56 \\
$yz$           & 16.57 & 17.06 & 17.07 & 15.78 & 15.32  & 3.28 & 4.22 & 3.62 & 3.12 & 3.07 \\
$3z^{2}-r^{2}$ & 17.27 & 17.06 & 19.03 & 17.07 & 15.95  & 3.29 & 3.62 & 4.75 & 3.62 & 3.08 \\
$zx$           & 16.57 & 15.78 & 17.07 & 17.06 & 15.32  & 3.28 & 3.12 & 3.62 & 4.22 & 3.07 \\
$x^{2}-y^{2}$  & 16.51 & 15.32 & 15.95 & 15.32 & 15.84  & 3.56 & 3.07 & 3.08 & 3.07 & 3.83 \\
\hline \hline \\ [-8pt]
               &      &      &     $J_{v}$     &      &               &      &      &      $J(0)$    &      &          \\ [+1pt]
\hline \\ [-8pt]
               & $xy$ & $yz$ & $3z^{2}-r^{2}$ & $zx$ & $x^{2}-y^{2}$ & $xy$ & $yz$ & $3z^{2}-r^{2}$ & $zx$ & $x^{2}-y^{2}$ \\ 
\hline \\ [-8pt] 
$xy$           &      & 0.66 & 0.79 & 0.66 & 0.34 &      & 0.57 & 0.69 & 0.57 & 0.32  \\
$yz$           & 0.66 &      & 0.46 & 0.57 & 0.61 & 0.57 &      & 0.42 & 0.49 & 0.53  \\
$3z^{2}-r^{2}$ & 0.79 & 0.46 &      & 0.46 & 0.74 & 0.69 & 0.42 &      & 0.42 & 0.62  \\
$zx$           & 0.58 & 0.57 & 0.46 &      & 0.61 & 0.57 & 0.49 & 0.42 &      & 0.53  \\
$x^{2}-y^{2}$  & 0.34 & 0.61 & 0.74 & 0.61 &      & 0.32 & 0.53 & 0.62 & 0.53 &       \\
\hline \hline \\ [-8pt]
               &      &      &    $v_{n}$     &       &             &       &      &      $V(0)$    &     &      \\ [+1pt]
\hline \\ [-8pt]
               & $xy$ & $yz$ & $3z^{2}-r^{2}$ & $zx$ & $x^{2}-y^{2}$ & $xy$ & $yz$ & $3z^{2}-r^{2}$ & $zx$ & $x^{2}-y^{2}$  \\ 
\hline \\ [-8pt] 
$xy$           & 5.32 & 5.23 & 5.21 & 5.23 & 5.23 & 1.22 & 1.20 & 1.19 & 1.20 & 1.22   \\
$yz$           & 5.23 & 5.13 & 5.12 & 5.15 & 5.14 & 1.20 & 1.19 & 1.18 & 1.21 & 1.20   \\
$3z^{2}-r^{2}$ & 5.21 & 5.12 & 5.10 & 5.12 & 5.13 & 1.19 & 1.18 & 1.16 & 1.18 & 1.19   \\
$zx$           & 5.23 & 5.15 & 5.12 & 5.13 & 5.14 & 1.20 & 1.21 & 1.18 & 1.19 & 1.20   \\
$x^{2}-y^{2}$  & 5.23 & 5.14 & 5.13 & 5.14 & 5.16 & 1.22 & 1.20 & 1.19 & 1.20 & 1.22   \\
\hline
\hline 
\end{tabular} 
\end{table*}

These results are understood from the transfer integrals and on-site energies. 
The transfer integrals, $t_{mn}(\bm{R})$, are summarized in Table \ref{t_FeSe}.
In the tables and this subsection, the symmetry of the $d$ orbitals is denoted as the number; $1$ for $xy$, $2$ for $yz$,
$3$ for $3z^{2}-r^{2}$, $4$ for $zx$, and $5$ for $x^{2}-y^{2}$ orbitals\cite{miyake2}.
The on-site energy listed in the column for $(\bm{R}_{x},\bm{R}_{y},\bm{R}_{z})=(0,0,0)$ and the nearest-neighbor transfer integral listed in the column for $(1/2,-1/2,0)$ are extracted in Figs. \ref{tonFeSe} and \ref{tnnFeSe}, respectively.

It is useful to discuss the behavior of $\mathcal{H}^{H}$ to understand the parameter values for $\mathcal{H}^{H\Delta}$.  In comparison to the LDA and $\mathcal{H}^{H\Delta}$ approximation, 
the on-site energy of the $x^2-y^2$ is lower in $\mathcal{H}^{H}$, and the transfer integral for $x^2-y^2$ 
(778 meV) is much larger than that in LDA (55 meV). 
Both of them widen the pDOS of the $x^2-y^2$ orbital. 
The on-site energy of the $3z^2-r^2$ is higher in $\mathcal{H}^{H}$, 
which results in reduction of the occupation number. 
The next-nearest transfer integrals $(\bm{R}_{x},\bm{R}_{y},\bm{R}_{z})=(1,0,0)$ 
are comparable with the nearest-neighbor ones in the LDA. 
Especially, the $(m,n)=(zx,zx)$ component, $t'^{\text{LDA}}_{44}$, is nearly the same as $t^{\text{LDA}}_{11}$, 
which makes the system frustrated.
These strong tendencies in $\mathcal{H}^{H}$ are substantially weakened in $\mathcal{H}^{H\Delta}$, but still a similar difference remains relative to the LDA result.

In the $\mathcal{H}^{H\Delta}$ case, the order of the on-site energies is partially changed from the $\mathcal{H}^{H}$ case. 
The crystal-field splitting is reduced and the transfer integrals change in both magnitude and order, 
which brings qualitative modification of the band dispersion. 
Compared to the LDA, the on-site energy of the $x^2-y^2$ is lower by 0.2 eV and the occupation number increases, while the $3z^2-r^2$ orbital is shifted upward by as large as 0.4 eV and the occupation number decreases.
As is the case with the LDA, the nearest-neighbor transfer integral for $(m,n)=(xy,xy)$, $t^{H\Delta}_{11}$ 
is the largest in magnitude. 
The value is $-460$ meV and is about $10$\% larger than that of the LDA ($-409$ meV).  
The $(m,n)=(zx,zx)$ transfer, $t'^{H\Delta}_{44}$, is about $30$\% smaller than $t^{H\Delta}_{11}$. 
The transfer integrals $t^{H\Delta}$ except for the $yz/zx$ orbitals become larger than those of the LDA, and 
the increase is substantial for the $3z^{2}-r^{2}$ and $x^{2}-y^{2}$.
Since the $x^{2}-y^{2}$ orbital is primarily responsible for the strong correlation effects with the Mott physics~\cite{misawa11,misawa12}, the increase in its transfer may cause some reduction of the correlation effects.


%
\begin{figure*}[htb]
\begin{center} 
\includegraphics[width=0.9\textwidth ]{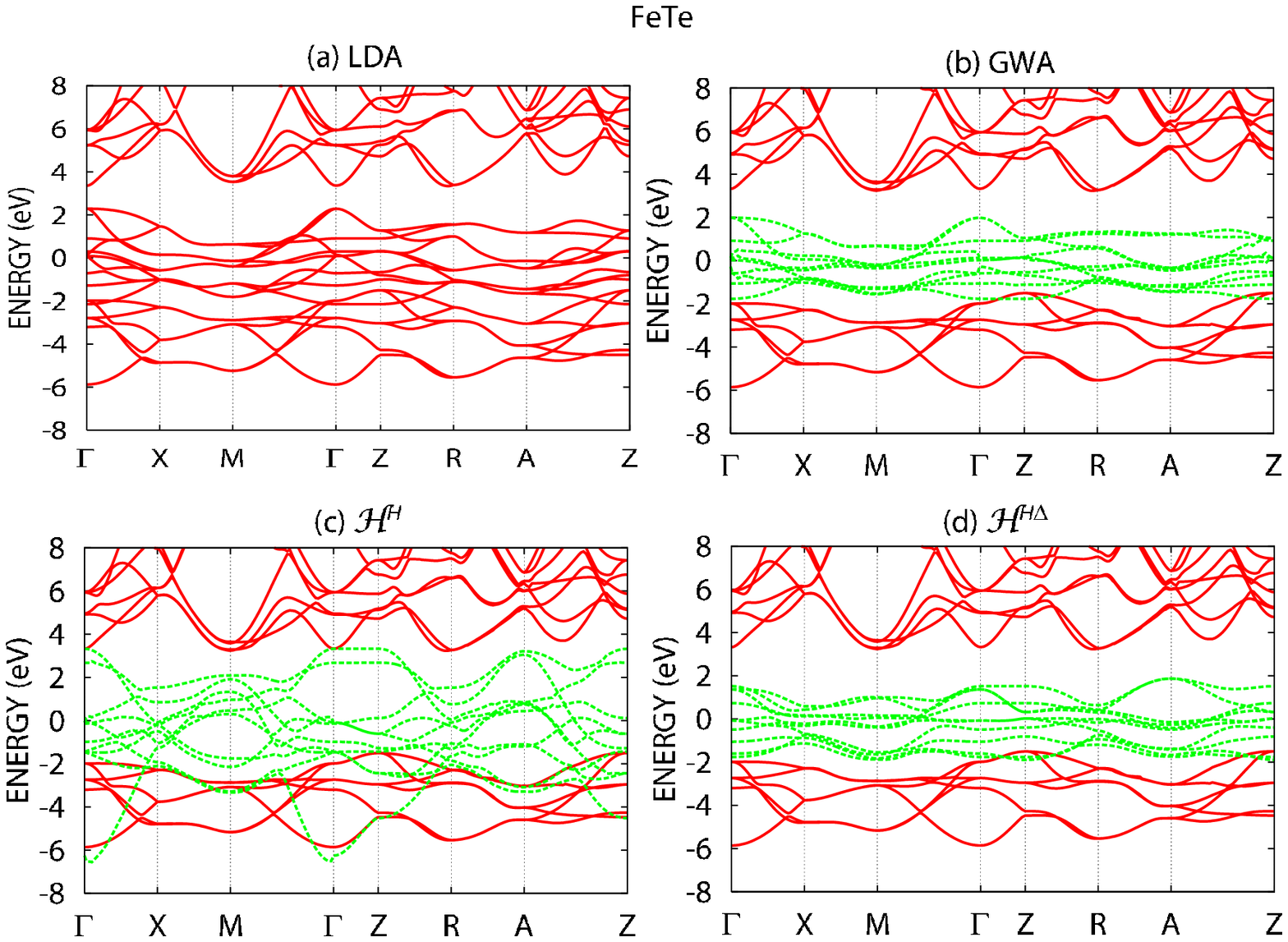} 
\end{center} 
\caption{(Color online) Electronic band structures of FeTe obtained from (a) LDA, (b) GWA, (c) Eq. (\ref{HGrW}) (${\mathcal H}^H$, excluding double counting in low-energy space)  and (d) Eq. (\ref{HGrWdelta}) (${\mathcal H}^{H\Delta}$, renormalizing the $\omega$-dependence of $W^{p}$ in addition to ${\mathcal H}^H$). The zero energy corresponds to the Fermi level. }
\label{bndsFeTe}
\end{figure*} 
\begin{figure*}[htb]
\begin{center} 
\includegraphics[width=0.9\textwidth ]{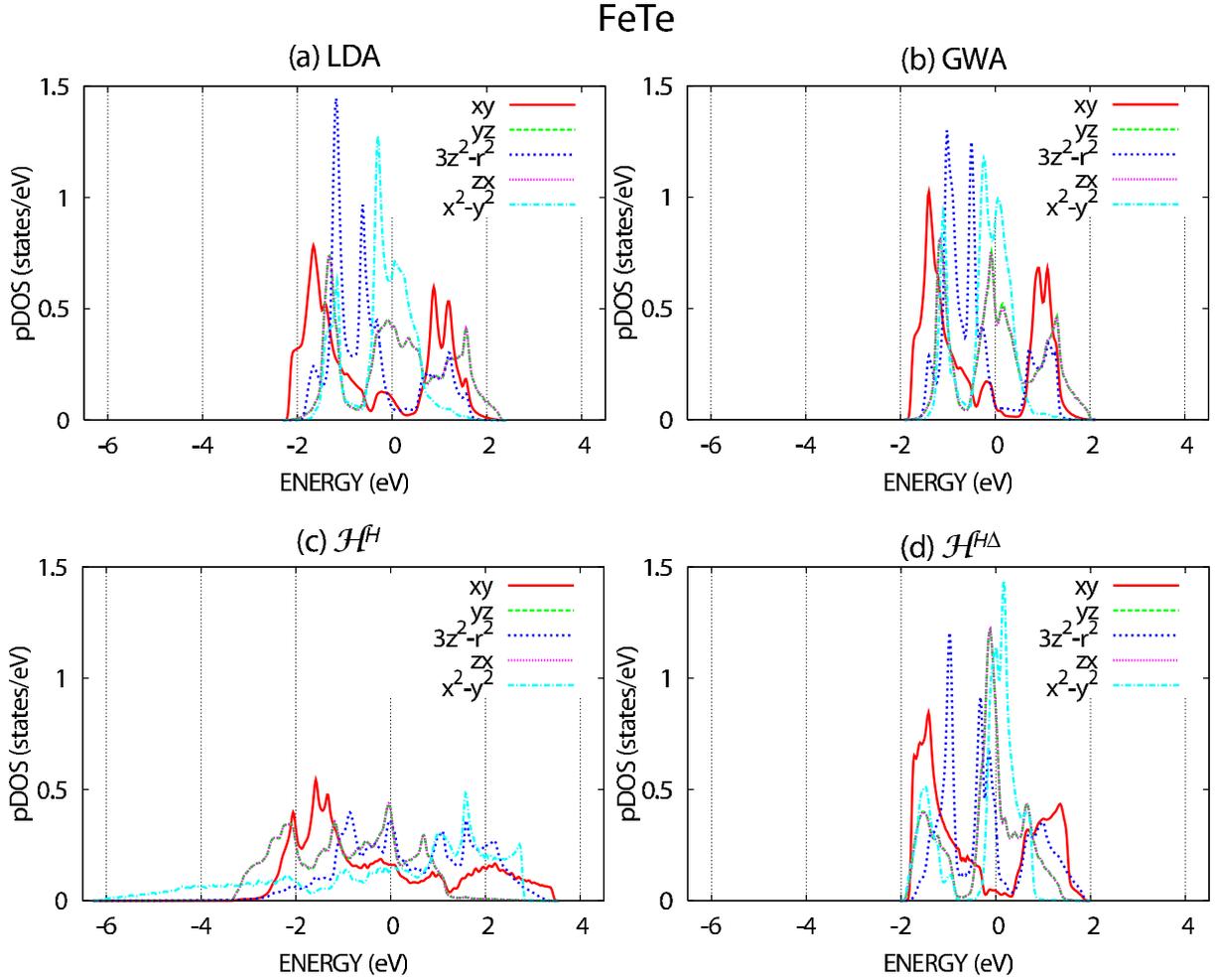} 
\end{center} 
\caption{(Color online) Partial densities of states for FeTe resolved by Wannier functions of Fe $3d$ orbitals calculated from (a)  LDA, (b) GWA, (c) Eq. (\ref{rG2}) (${\mathcal H}^H$, excluding double counting in low-energy space) and (d) Eq. (\ref{rdelG2}) (${\mathcal H}^{H\Delta}$, renormalizing the $\omega$-dependence of $W^{p}$ in addition to ${\mathcal H}^H$). The zero energy corresponds to the Fermi level. }
\label{DOSFeTe}
\end{figure*} 

\begin{table}[tb] 
\caption{Occupation number of Wannier functions of FeTe, where total occupancy is $6$.
} 
\
\begin{tabular}{c|ccccc}
\hline \hline \\ [-8pt]  
Occ. Num.               & $xy$ & $yz$ & $3z^{2}-r^{2}$ & $zx$ & $x^{2}-y^{2}$   \\ [+1pt]
\hline \\ [-8pt] 
LDA                     & 1.27 & 0.96 & 1.58 & 0.96 & 1.23    \\ 
GWA                     & 1.25 & 0.99 & 1.54 & 0.99 & 1.24    \\ 
$\mathcal{H}^{H}$       & 1.28 & 1.52 & 0.86 & 1.52 & 0.82    \\ 
$\mathcal{H}^{H\Delta}$ & 1.21 & 1.25 & 1.35 & 1.25 & 0.94    \\ 
\hline \hline 

\end{tabular}
\label{Occ_FeTe} 
\end{table} 

\begin{figure}[ptb]
\begin{center} 
\includegraphics[width=0.4\textwidth ]{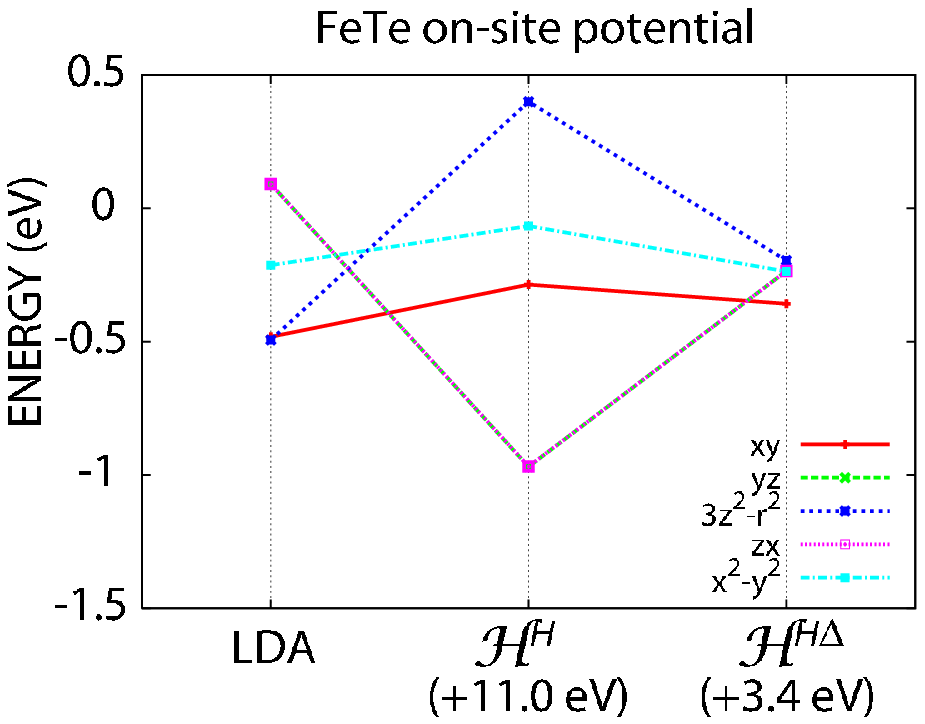} 
\end{center} 
\caption{On-site potential of Wannier orbitals for FeTe.
Figure in the bracket is the value of the constant shift from the Fermi level of the high-energy space.
}
\label{tonFeTe}
\end{figure} 

We next show results for FeTe.
Figure \ref{bndsFeTe} (a) shows the band structure of FeTe in the LDA.
The ten-fold Fe $3d$ bands partially entangle with the Te $5p$ valence bands, while the low-energy bands are similar to that of FeSe.  
In fact, the density of states of FeTe for the Fe $5d$ bands shown in Figure \ref{DOSFeTe} (a) reveals that 
the $x^{2}-y^{2}$ orbital has the largest density of states at the Fermi level followed by the $yz$/$zx$ orbitals similarly to the case of FeSe discussed above. In case of FeTe, these two orbitals even have peaks near the Fermi level.
In Table \ref{Occ_FeTe}, we see that the occupation number is also similar to that of FeSe.

Figures \ref{bndsFeTe} (b) and \ref{DOSFeTe} (b) show the band structure and the density of states of FeTe in the GWA, respectively, where we disentangle the Fe $3d$ Wannier bands from the whole KS-band structure\cite{imada,miyake09} and calculate the self-energy only for them. 
The distribution of the density of states and the occupation number are again nearly the same as those of the LDA and similar to those of FeSe (see Table \ref{Occ_FeTe}).

We now show in Figs. \ref{bndsFeTe} (c) and (d) the band structure of FeTe after including the self-energy effect described by ${\cal H}^H$ introduced in Eq. (\ref{HGrW}) and ${\cal H}^{H\Delta}$ in Eq. (\ref{HGrWdelta}), respectively, excluding the double counting of the electron correlation.
As is the case with FeSe, the band structures are substantially different from that of the LDA.
The corresponding densities of states are also shown in Figs.~\ref{DOSFeTe} (c) and (d).
We summarize the transfer integrals of FeTe without the double counting of the electron correlation in Table \ref{t_FeTe}.

Now we focus on the effective model parameters obtained from ${\cal H}^{H\Delta}$ in Eq. (\ref{HGrWdelta}).  
The transfer integrals $t^{H\Delta}$ except for the $yz/zx$ orbitals become larger than those of the LDA, and 
the increase is substantial for the $3z^{2}-r^{2}$ and $x^{2}-y^{2}$, similarly to FeSe. 

Geometrical frustration effects measured by the ratio between the next-nearest-neighbor ($t'$) to the nearest-neighbor ($t$) transfers also show a tendency similar to the case of FeSe: The frustration $|t'/t|$ is remarkably suppressed from 30/71 to 56/279 for the diagonal transfer between two $x^{2}-y^{2}$ orbitals.
Although this reduction looks dominating the magnetic stability, the frustration for the $3z^{2}-r^{2}$ orbital increases because $t$ and $t'$ both become more than twice of the LDA parameters and have similar amplitudes ($t=-99$ meV and $t'=-154$ eV).
All of these corrections from the LDA results are qualitatively similar to the case of FeSe.
The increase of $t'$ between the $3z^{2}-r^{2}$ and $x^{2}-y^{2}$ orbitals to $66$ meV is also noticeable, which is different from the case of FeSe.

The stripe order is in general stabilized when $t'$ becomes comparable to $t$ in amplitude, while the reduction of $|t'/t|$ may stabilize longer-period antiferromagnetic order such as up-up-down-down structure~\cite{mizusaki} as is observed as the double stripe order in FeTe before the simple staggered G-type antiferromagnetic order becomes stabilized.  The previous solution for the {\it ab initio} low-energy models by the variational Monte Carlo method\cite{misawa12} based on the LDA band structure has shown nearly degenerate ground states of the stripe and double stripe structure. By considering the overall reduction of the frustration particularly for the pair at the $x^{2}-y^{2}$ orbitals, it is an interesting issue to study whether the double stripe phase becomes the unique ground state in the present {\it ab initio} model as in the experimental observation.

\begin{table*}[ptb] 
\caption{Transfer integral and its components for the $3d$ orbitals of the Fe sites in FeTe, $t_{mn}(R_x, R_y, R_z)$, 
where $m$ and $n$ denote symmetries of $3d$ orbitals. Units are given in meV.  
$t^{\text{LDA}}$ is the expectation value of the KS-Hamiltonian for the Wannier function : $^{\text{LDA}}=\langle \phi^{L}| \mathcal{H}^{\text{LDA}} |\phi^{L} \rangle$ (Eq. (\ref{tLDA})).   
$t^{H}$ is the static transfer integral without double-counting : $t^{H}=\langle \phi^{L}| \mathcal{H}^{\text{LDA}}+Z^{H}(-V^{\text{xc}}+\text{Re}\Sigma ^{H}) |\phi^{L} \rangle$  (Eq. (\ref{HGrW})). 
$t^{H\Delta}$ is the static transfer integral including the correction of the $\omega $-dependence of $U$ : $t^{H\Delta}=\langle \phi^{L}|\mathcal{H}^{\text{LDA}}+Z^{H\Delta}(-V^{\text{xc}}+\text{Re}(\Sigma^{H}+\Delta \Sigma ^{L}))|\phi^{L} \rangle$ (Eq. (\ref{HGrWdelta})).
Figure in the bracket in [0,0,0] is the value of the constant shift from the Fermi level of the high-energy space.
} 
\
\begin{tabular}{c|rrrrrrr|rrr} 
\hline \hline \\ [-4pt]
  $t^{\text{LDA}}$  \\ [+2pt] 
\hline \\ [-4pt]
$(m, n)$ $\backslash$ $\bm{R}$ 
& \big[$0,0,0$\big] 
& \big[$\frac{1}{2},-\frac{1}{2},0$\big] 
& \big[$1,0,0$\big] 
& \big[$1,-1,0$\big] 
& \big[$\frac{3}{2},-\frac{1}{2},0$\big]
& \big[$0,0,\frac{c}{a}$\big] 
& \big[$\frac{1}{2},-\frac{1}{2},\frac{c}{a}$\big]
& $\sigma_{y}$
& $I$
& $\sigma^{L}$ \\ [+4pt]
\hline \\ [-8pt]             
$(xy,xy)$                    & $-$482 & $-$381 &      9 & $-$44 &  $-$1 & $-$31 &    15 & $+$ & $+$ &      $+$ \\ 
$(xy,yz)$                    &      0 &    243 &    110 &  $-$3 &     0 &     0 & $-$11 & $+$ & $-$ & $-$(1,4) \\
$(xy,3z^{2}-r^{2})$          &      0 & $-$330 &      0 &    35 &     0 &     0 &    22 & $-$ & $+$ &      $+$ \\ 
$(xy,zx)$                    &      0 &    243 &      0 &  $-$3 &    72 &     0 &     2 & $-$ & $-$ & $-$(1,2) \\
$(xy,x^{2}-y^{2})$           &      0 &      0 &      0 &     0 &     0 &     0 &     6 & $-$ & $+$ &      $-$ \\ 
$(yz,yz)$                    &     91 &    163 &     88 & $-$17 & $-$24 &    15 &    44 & $+$ & $+$ &    (4,4) \\ 
$(yz,3z^{2}-r^{2})$          &      0 & $-$124 &      0 &    20 &    16 &     0 &    10 & $-$ & $-$ & $-$(4,3) \\ 
$(yz,zx)$                    &      0 &    101 &      0 & $-$24 &     0 &     0 &    12 & $-$ & $+$ &    (4,2) \\ 
$(yz,x^{2}-y^{2})$           &      0 &    176 &      0 &     1 & $-$19 &     0 &    19 & $-$ & $-$ &    (4,5) \\ 
$(3z^{2}-r^{2},3z^{2}-r^{2})$& $-$494 &  $-$59 &  $-$85 &  $-$4 &    18 & $-$66 & $-$21 & $+$ & $+$ &      $+$ \\ 
$(3z^{2}-r^{2},zx)$          &      0 &    124 &    177 & $-$20 &     0 &     0 & $-$28 & $+$ & $-$ & $-$(3,2) \\ 
$(3z^{2}-r^{2},x^{2}-y^{2})$ &      0 &      0 &  $-$19 &     0 & $-$17 &    31 & $-$26 & $+$ & $+$ &      $-$ \\ 
$(zx,zx)$                    &     91 &    163 &    394 & $-$17 &    84 &    15 &     9 & $+$ & $+$ &    (2,2) \\ 
$(zx,x^{2}-y^{2})$           &      0 & $-$176 &    142 &  $-$1 &     0 &     0 &    23 & $+$ & $-$ &    (2,5) \\ 
$(x^{2}-y^{2},x^{2}-y^{2})$  & $-$213 &     71 &     30 &     1 &    27 &    10 & $-$16 & $+$ & $+$ &      $+$ \\ 
\hline \hline 
  $t^{H}$    \\ [+2pt] 
\hline \\ [-4pt]
$(m, n)$ $\backslash$ $\bm{R}$ 
& \big[$0,0,0$\big] 
& \big[$\frac{1}{2},-\frac{1}{2},0$\big] 
& \big[$1,0,0$\big] 
& \big[$1,-1,0$\big] 
& \big[$\frac{3}{2},-\frac{1}{2},0$\big]
& \big[$0,0,\frac{c}{a}$\big] 
& \big[$\frac{1}{2},-\frac{1}{2},\frac{c}{a}$\big]
& $\sigma_{y}$
& $I$
& $\sigma^{L}$ \\ [+4pt]
\hline \\ [-8pt]             
$(xy,xy)$                    &  $-$286(+11024) & $-$606 &    205 &   126 &      1 & $-$105 &    20 & $+$ & $+$ &      $+$ \\ 
$(xy,yz)$                    &       0 & $-$136 &     45 & $-$18 &      0 &      1 &     5 & $+$ & $-$ & $-$(1,4) \\
$(xy,3z^{2}-r^{2})$          &       0 & $-$396 &      1 &    27 &      0 &   $-$3 &    42 & $-$ & $+$ &      $+$ \\ 
$(xy,zx)$                    &       0 & $-$135 &   $-$1 & $-$20 & $-$172 &      0 & $-$17 & $-$ & $-$ & $-$(1,2) \\
$(xy,x^{2}-y^{2})$           &    $-$1 &      0 &   $-$1 &     0 &      0 &      0 & $-$18 & $-$ & $+$ &      $-$ \\ 
$(yz,yz)$                    &  $-$969(+11024) & $-$225 &    102 & $-$48 &  $-$92 &     41 &    36 & $+$ & $+$ &    (4,4) \\ 
$(yz,3z^{2}-r^{2})$          &       0 &     11 &      1 &     3 &     25 &      1 & $-$21 & $-$ & $-$ & $-$(4,3) \\ 
$(yz,zx)$                    &       0 &    455 &      0 & $-$53 &      2 &      0 &  $-$5 & $-$ & $+$ &    (4,2) \\ 
$(yz,x^{2}-y^{2})$           &       0 &     30 &   $-$1 & $-$12 &     42 &      0 & $-$12 & $-$ & $-$ &    (4,5) \\ 
$(3z^{2}-r^{2},3z^{2}-r^{2})$&     400(+11024) & $-$252 & $-$285 &    66 &    136 &     59 &    16 & $+$ & $+$ &      $+$ \\ 
$(3z^{2}-r^{2},zx)$          &       1 &  $-$11 &    148 &  $-$4 &      0 &   $-$1 &     7 & $+$ & $-$ & $-$(3,2) \\ 
$(3z^{2}-r^{2},x^{2}-y^{2})$ &       0 &      0 &    283 &     0 & $-$129 &  $-$47 &    26 & $+$ & $+$ &      $-$ \\ 
$(zx,zx)$                    &  $-$968(+11024) & $-$225 &  $-$50 & $-$48 &    148 &     41 &    14 & $+$ & $+$ &    (2,2) \\ 
$(zx,x^{2}-y^{2})$           &       0 &  $-$28 &    185 &    10 &      1 &      0 & $-$15 & $+$ & $-$ &    (2,5) \\ 
$(x^{2}-y^{2},x^{2}-y^{2})$  &   $-$66(+11024) &    983 & $-$288 & $-$11 &     41 &  $-$52 &    41 & $+$ & $+$ &      $+$ \\ 
\hline \hline 
  $t^{H\Delta}$    \\ [+2pt] 
\hline \\ [-4pt]
$(m, n)$ $\backslash$ $\bm{R}$ 
& \big[$0,0,0$\big] 
& \big[$\frac{1}{2},-\frac{1}{2},0$\big] 
& \big[$1,0,0$\big] 
& \big[$1,-1,0$\big] 
& \big[$\frac{3}{2},-\frac{1}{2},0$\big]
& \big[$0,0,\frac{c}{a}$\big] 
& \big[$\frac{1}{2},-\frac{1}{2},\frac{c}{a}$\big]
& $\sigma_{y}$
& $I$
& $\sigma^{L}$ \\ [+4pt]
\hline \\ [-8pt]             
$(xy,xy)$                    & $-$358(+3447) & $-$421 &     61 &     2 &     1 & $-$57 &    17 & $+$ & $+$ &      $+$ \\ 
$(xy,yz)$                    &   $-$1 &    135 &     75 & $-$11 &     0 &     0 &  $-$4 & $+$ & $-$ & $-$(1,4) \\
$(xy,3z^{2}-r^{2})$          &      0 & $-$331 &      0 &    24 &     1 &  $-$1 &    30 & $-$ & $+$ &      $+$ \\ 
$(xy,zx)$                    &      1 &    135 &   $-$1 & $-$12 &  $-$1 &     0 &  $-$7 & $-$ & $-$ & $-$(1,2) \\
$(xy,x^{2}-y^{2})$           &   $-$1 &      0 &      0 &     0 &     0 &     0 &  $-$1 & $-$ & $+$ &      $-$ \\ 
$(yz,yz)$                    & $-$235(+3447) &     77 &    107 & $-$26 & $-$55 &    26 &    49 & $+$ & $+$ &    (4,4) \\ 
$(yz,3z^{2}-r^{2})$          &   $-$1 &  $-$80 &      0 &    23 &    19 &     0 &     1 & $-$ & $-$ & $-$(4,3) \\ 
$(yz,zx)$                    &      0 &    171 &      0 & $-$25 &     1 &     0 &     7 & $-$ & $+$ &    (4,2) \\ 
$(yz,x^{2}-y^{2})$           &      0 &    120 &   $-$1 &  $-$2 &  $-$6 &     0 &     5 & $-$ & $-$ &    (4,5) \\ 
$(3z^{2}-r^{2},3z^{2}-r^{2})$& $-$196(+3447) &  $-$99 & $-$154 &    31 &    44 & $-$36 &  $-$8 & $+$ & $+$ &      $+$ \\ 
$(3z^{2}-r^{2},zx)$          &      1 &     80 &    155 & $-$23 &     0 &  $-$1 & $-$19 & $+$ & $-$ & $-$(3,2) \\ 
$(3z^{2}-r^{2},x^{2}-y^{2})$ &      0 &      0 &     66 &     0 & $-$45 &     9 &  $-$8 & $+$ & $+$ &      $-$ \\ 
$(zx,zx)$                    & $-$235(+3447) &     77 &    268 & $-$26 &   117 &    26 &     8 & $+$ & $+$ &    (2,2) \\ 
$(zx,x^{2}-y^{2})$           &      0 & $-$120 &    134 &     1 &     0 &     0 &     7 & $+$ & $-$ &    (2,5) \\ 
$(x^{2}-y^{2},x^{2}-y^{2})$  & $-$237(+3447) &    279 &  $-$56 & $-$11 &    46 &  $-$6 &  $-$5 & $+$ & $+$ &      $+$ \\ 
\hline \hline 
\end{tabular}
\label{t_FeTe} 
\end{table*} 

\begin{table*}[htb] 
\caption{
Bare and effective Coulomb interactions between two electrons for all the combinations of Fe $3d$ orbitals in FeTe (in eV).
Here, $v$ and $J_{v}$ represent the bare on-site and exchange Coulomb interactions, respectively. 
The static limit ($\omega\rightarrow 0$) of the effective on-site and exchange Coulomb interactions are denoted by $U(0)$ and $J(0)$, while
$v_{n}$ and $V(0)$ represent the  bare and effective nearest-neighbor Coulomb interactions, respectively. 
}
\ 
\label{W_FeTe} 
\begin{tabular}{c|ccccc|ccccc} 
\hline \hline \\ [-8pt]
FeTe           &      &      &     $v$     &      &               &      &      &      $U(0)$    &      &          \\ [+1pt]
\hline \\ [-8pt]
               & $xy$ & $yz$ & $3z^{2}-r^{2}$ & $zx$ & $x^{2}-y^{2}$ & $xy$ & $yz$ & $3z^{2}-r^{2}$ & $zx$ & $x^{2}-y^{2}$ \\ 
\hline \\ [-8pt] 
$xy$           & 17.17 & 15.01 & 16.28 & 15.01 & 16.32  & 3.49 & 2.31 & 2.37 & 2.31 & 2.82\\
$yz$           & 15.01 & 15.19 & 15.80 & 14.08 & 14.87  & 2.31 & 3.05 & 2.62 & 2.14 & 2.30\\
$3z^{2}-r^{2}$ & 16.28 & 15.80 & 18.33 & 15.80 & 16.13  & 2.37 & 2.62 & 3.74 & 2.62 & 2.36\\
$zx$           & 15.01 & 14.08 & 15.80 & 15.19 & 14.87  & 2.31 & 2.14 & 2.62 & 3.05 & 2.30\\
$x^{2}-y^{2}$  & 16.32 & 14.87 & 16.13 & 14.87 & 16.80  & 2.82 & 2.30 & 2.36 & 2.30 & 3.39\\
\hline \hline \\ [-8pt]
               &      &      &     $J_{v}$     &      &               &      &      &      $J(0)$    &      &          \\ [+1pt]
\hline \\ [-8pt]
               & $xy$ & $yz$ & $3z^{2}-r^{2}$ & $zx$ & $x^{2}-y^{2}$ & $xy$ & $yz$ & $3z^{2}-r^{2}$ & $zx$ & $x^{2}-y^{2}$ \\ 
\hline \\ [-8pt] 
$xy$           &      & 0.58 & 0.73 & 0.58 & 0.33 &      & 0.48 & 0.62 & 0.48 & 0.31 \\
$yz$           & 0.58 &      & 0.41 & 0.48 & 0.58 & 0.48 &      & 0.37 & 0.39 & 0.49 \\
$3z^{2}-r^{2}$ & 0.73 & 0.41 &      & 0.41 & 0.75 & 0.62 & 0.37 &      & 0.37 & 0.62 \\
$zx$           & 0.58 & 0.48 & 0.41 &      & 0.58 & 0.48 & 0.39 & 0.37 &      & 0.49 \\
$x^{2}-y^{2}$  & 0.33 & 0.58 & 0.75 & 0.58 &      & 0.31 & 0.49 & 0.62 & 0.49 &      \\
\hline \hline \\ [-8pt]
               &      &      &    $v_{n}$     &       &             &       &      &      $V(0)$    &     &      \\ [+1pt]
\hline \\ [-8pt]
               & $xy$ & $yz$ & $3z^{2}-r^{2}$ & $zx$ & $x^{2}-y^{2}$ & $xy$ & $yz$ & $3z^{2}-r^{2}$ & $zx$ & $x^{2}-y^{2}$  \\ 
\hline \\ [-8pt] 
$xy$           & 5.12 & 5.01 & 5.04 & 5.01 & 5.11 & 0.97 & 0.96 & 0.95 & 0.96 & 0.97   \\
$yz$           & 5.01 & 4.91 & 4.95 & 4.93 & 5.00 & 0.96 & 0.95 & 0.94 & 0.96 & 0.95   \\
$3z^{2}-r^{2}$ & 5.04 & 4.95 & 4.96 & 4.95 & 5.04 & 0.95 & 0.94 & 0.93 & 0.94 & 0.94   \\
$zx$           & 5.01 & 4.93 & 4.95 & 4.91 & 5.00 & 0.96 & 0.96 & 0.94 & 0.95 & 0.95   \\
$x^{2}-y^{2}$  & 5.11 & 5.00 & 5.04 & 5.00 & 5.11 & 0.97 & 0.95 & 0.94 & 0.95 & 0.98   \\
\hline
\hline 
\end{tabular} 
\end{table*}

\begin{figure*}[htb]
\begin{center} 
\includegraphics[width=0.95\textwidth ]{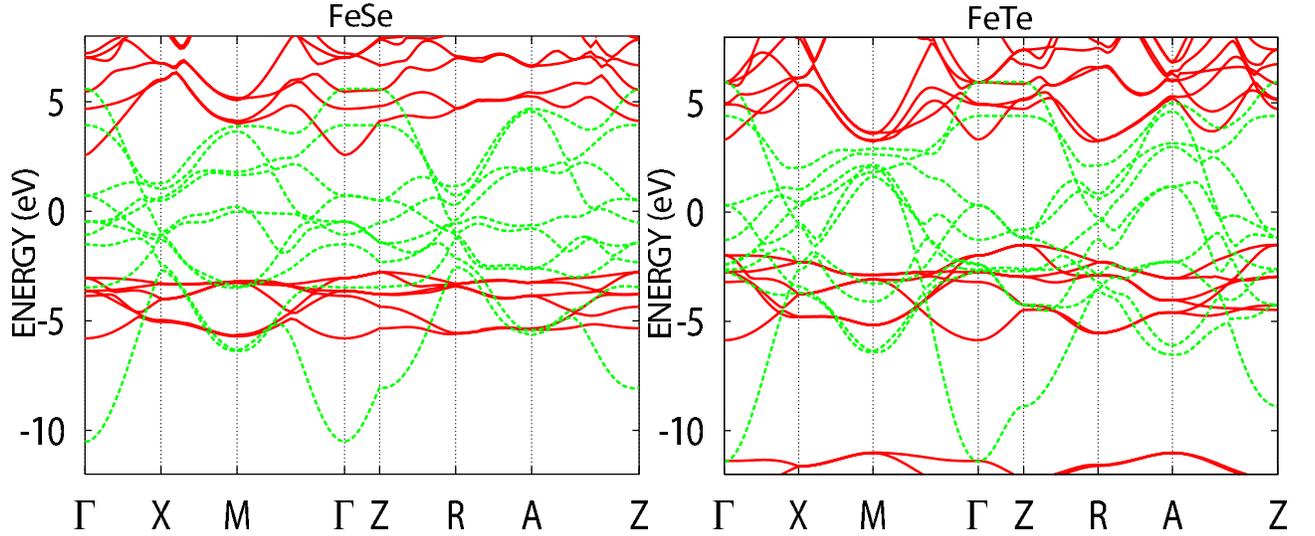} 
\end{center} 
\caption{(Color online) Electronic band structures calculated without exchange-correlation potential of FeSe and FeTe. The zero energy corresponds to the Fermi level. }
\label{bndsFeSeandFeTeFock0}
\end{figure*} 
\begin{figure*}[htb]
\begin{center} 
\includegraphics[width=0.95\textwidth ]{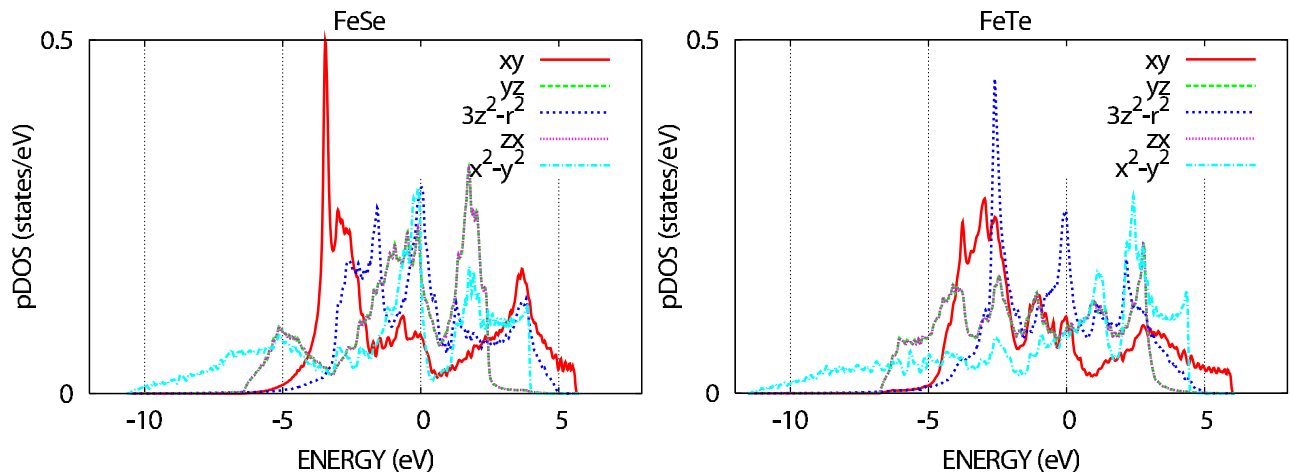} 
\end{center} 
\caption{(Color online) Partial densities of states calculated without  exchange-correlation potential for FeSe and FeTe resolved by Wannier functions of Fe $3d$ orbitals. }
\label{DOSFeSeandFeTeFock0}
\end{figure*} 

\begin{table*}[htb] 
\caption{Transfer integral between Fe $3d$ Wannier orbitals for FeSe and FeTe, $t_{mn}(R_x, R_y, R_z)$, 
where $m$ and $n$ denote symmetries of $3d$ orbitals. Here, Fock terms are not included. Units are given in meV.  
Transfer integral calculated without exchange-correlation potential is given as $t^{0}=\langle \phi^{L}| \mathcal{H}^{\text{LDA}}-V^{\text{xc}}|\phi^{L} \rangle$.
Figure in the bracket in [0,0,0] is the value of the constant shift from the Fermi level of the high-energy space.
} 
\
\begin{tabular}{c|rrrrrrr|rrr} 
\hline \hline \\ [-4pt]
  FeSe  \\ [+2pt] 
\hline \\ [-4pt]
$(m, n)$ $\backslash$ $\bm{R}$ 
& \big[$0,0,0$\big] 
& \big[$\frac{1}{2},-\frac{1}{2},0$\big] 
& \big[$1,0,0$\big] 
& \big[$1,-1,0$\big] 
& \big[$\frac{3}{2},-\frac{1}{2},0$\big]
& \big[$0,0,\frac{c}{a}$\big] 
& \big[$\frac{1}{2},-\frac{1}{2},\frac{c}{a}$\big]
& $\sigma_{y}$
& $I$
& $\sigma^{L}$ \\ [+4pt]
\hline \\ [-8pt]             
$(xy,xy)$                    &  $-$347(+27908) & $-$1381 &    167 &    185 &     12 & $-$126 &    22 & $+$ & $+$ &      $+$ \\ 
$(xy,yz)$                    &       0 &     113 &    200 &  $-$86 &      3 &      0 &    13 & $+$ & $-$ & $-$(1,4) \\
$(xy,3z^{2}-r^{2})$          &       0 &  $-$859 &      0 &     91 &      0 &      0 &    20 & $-$ & $+$ &      $+$ \\ 
$(xy,zx)$                    &       0 &     113 &      0 &  $-$86 & $-$209 &      0 & $-$20 & $-$ & $-$ & $-$(1,2) \\
$(xy,x^{2}-y^{2})$           &       0 &       0 &      0 &      0 &      0 &      0 & $-$35 & $-$ & $+$ &      $-$ \\ 
$(yz,yz)$                    &  $-$749(+27908) &  $-$261 &    347 &  $-$90 &  $-$86 &     33 &    78 & $+$ & $+$ &    (4,4) \\ 
$(yz,3z^{2}-r^{2})$          &       0 &   $-$74 &      0 &     75 &      5 &      0 & $-$18 & $-$ & $-$ & $-$(4,3) \\ 
$(yz,zx)$                    &       0 &     951 &      0 & $-$134 &      0 &      0 &    27 & $-$ & $+$ &    (4,2) \\ 
$(yz,x^{2}-y^{2})$           &       0 &     120 &      0 &  $-$19 &     83 &      0 &     5 & $-$ & $-$ &    (4,5) \\ 
$(3z^{2}-r^{2},3z^{2}-r^{2})$&   $-$59(+27908) &  $-$313 & $-$391 &     44 &    147 &     46 &    53 & $+$ & $+$ &      $+$ \\ 
$(3z^{2}-r^{2},zx)$          &       0 &      74 &    138 &  $-$75 &      0 &      0 &    27 & $+$ & $-$ & $-$(3,2) \\ 
$(3z^{2}-r^{2},x^{2}-y^{2})$ &       0 &       0 &    362 &      0 & $-$117 &  $-$95 &    65 & $+$ & $+$ &      $-$ \\ 
$(zx,zx)$                    &  $-$749(+27908) &  $-$261 &     78 &  $-$90 &    195 &     33 &     2 & $+$ & $+$ &    (2,2) \\ 
$(zx,x^{2}-y^{2})$           &       0 &  $-$120 &    483 &     19 &      0 &      0 & $-$22 & $+$ & $-$ &    (2,5) \\ 
$(x^{2}-y^{2},x^{2}-y^{2})$  & $-$1301(+27908) &    1601 & $-$245  & $-$34 &  $-$26 &  $-$94 &    83 & $+$ & $+$ &      $+$ \\ 
\hline \hline 
  FeTe    \\ [+2pt] 
\hline \\ [-4pt]
$(m, n)$ $\backslash$ $\bm{R}$ 
& \big[$0,0,0$\big] 
& \big[$\frac{1}{2},-\frac{1}{2},0$\big] 
& \big[$1,0,0$\big] 
& \big[$1,-1,0$\big] 
& \big[$\frac{3}{2},-\frac{1}{2},0$\big]
& \big[$0,0,\frac{c}{a}$\big] 
& \big[$\frac{1}{2},-\frac{1}{2},\frac{c}{a}$\big]
& $\sigma_{y}$
& $I$
& $\sigma^{L}$ \\ [+4pt]
\hline \\ [-8pt]             
$(xy,xy)$                    &  $-$838(+27563) & $-$1048 &    400 &    216 &  $-$45 & $-$208 &    47 & $+$ & $+$ &      $+$ \\ 
$(xy,yz)$                    &       0 &  $-$195 &    111 &  $-$65 &      0 &      0 &    20 & $+$ & $-$ & $-$(1,4) \\
$(xy,3z^{2}-r^{2})$          &       0 &  $-$640 &      0 &     80 &   $-$1 &   $-$1 &    88 & $-$ & $+$ &      $+$ \\ 
$(xy,zx)$                    &       0 &  $-$194 &      0 &  $-$65 & $-$345 &      0 & $-$30 & $-$ & $-$ & $-$(1,2) \\
$(xy,x^{2}-y^{2})$           &    $-$1 &       0 &      0 &      0 &      0 &      0 & $-$20 & $-$ & $+$ &      $-$ \\ 
$(yz,yz)$                    & $-$1439(+27563) &  $-$421 &    266 &  $-$89 & $-$144 &    116 &    66 & $+$ & $+$ &    (4,4) \\ 
$(yz,3z^{2}-r^{2})$          &       2 &   $-$53 &      0 &     51 &     37 &      1 & $-$62 & $-$ & $-$ & $-$(4,3) \\ 
$(yz,zx)$                    &       0 &    1123 &   $-$1 & $-$134 &      1 &      0 & $-$19 & $-$ & $+$ &    (4,2) \\ 
$(yz,x^{2}-y^{2})$           &       0 &   $-$52 &   $-$1 &  $-$15 &     70 &      0 & $-$39 & $-$ & $-$ &    (4,5) \\ 
$(3z^{2}-r^{2},3z^{2}-r^{2})$&  $-$324(+27563) &  $-$387 & $-$473 &     98 &    223 &     17 &    15 & $+$ & $+$ &      $+$ \\ 
$(3z^{2}-r^{2},zx)$          &    $-$2 &      51 &     91 &  $-$51 &   $-$1 &   $-$1 & $-$20 & $+$ & $-$ & $-$(3,2) \\ 
$(3z^{2}-r^{2},x^{2}-y^{2})$ &       0 &       0 &    536 &      0 & $-$214 &  $-$25 &    24 & $+$ & $+$ &      $-$ \\ 
$(zx,zx)$                    & $-$1439(+27563) &  $-$421 & $-$303 &  $-$89 &    271 &    116 &    30 & $+$ & $+$ &    (2,2) \\ 
$(zx,x^{2}-y^{2})$           &       0 &      53 &    230 &     14 &      0 &      1 & $-$12 & $+$ & $-$ &    (2,5) \\ 
$(x^{2}-y^{2},x^{2}-y^{2})$  &  $-$661(+27563) &    1773 & $-$552 &   $-$1 &     74 &  $-$41 &    47 & $+$ & $+$ &      $+$ \\ 
\hline \hline 
\end{tabular}
\label{t_Fock0} 
\end{table*}

The effective Coulomb interactions for FeTe obtained from the cRPA are listed in Table \ref{W_FeTe}. The on-site $U$, the exchange $J$ and the nearest-neighbor diagonal Coulomb interaction $V$ are very similar to those obtained previously~\cite{miyake2} by using ${\cal H}^{\rm LDA}$ (namely, the LDA
band structure) and the differences are all within 10 \% and mostly below 0.2 eV.  
Similarly to the previous results~\cite{miyake2}, the overall correlation amplitudes are slightly smaller than those of FeSe.

A sharp contrast emerges in the comparison of the {\it ab initio} models for FeSe and FeTe:
The contrast is found in orbital level shifts (or the on-site potential shifts) of the $x^{2}-y^{2}$ orbital when we compare with the LDA band structure, as we see in the comparison between Figs. \ref{tonFeSe} and \ref{tonFeTe}.
For the present FeTe {\it ab initio} model, although the relative shift of the $x^{2}-y^{2}$ orbital level is not remarkable in Fig.\ref{tonFeTe}, the occupation number of the $x^2-y^2$ orbital is reduced from the LDA value 1.23 to 0.94 as we list in Table~\ref{Occ_FeTe}. 
On the contrary, FeSe shows a completely opposite behavior, where the occupation number of the $x^2-y^2$ orbital increases from the LDA value 1.11 to 1.35, because of a substantial downward shift of the $x^2-y^2$ orbital level relative to other orbitals.

We now elucidate the origin of this conspicuous downward shift of the $x^2-y^2$ orbital of FeSe. We find that this downward shift is even stronger in $t^{H}$ (actually the $x^2-y^2$ orbital becomes more than 0.5 eV lower than any other Fe 3$d$ orbital as one sees in Table \ref{t_FeSe}. 
In contrast, the occupation number of the $xy$ orbital, which has the band-insulator-like pDOS, is almost unchanged from that of the LDA ($\sim 6/5=1.2$). We then show the band structure, the density of states, and the transfer integral 
calculated without the exchange-correlation potential, $t^{0}=\langle \phi^{L}| \mathcal{H}^{\text{LDA}}-V^{\text{xc}}|\phi^{L} \rangle$, in Figs. \ref{bndsFeSeandFeTeFock0} and \ref{DOSFeSeandFeTeFock0}, and Table \ref{t_Fock0}, respectively.
Overall behavior is nearly the same as that of ``$t^{H}$" except for the uniform reduction of the band width.
The total width of the Fe $3d$ bands in ``$t^{0}$" is about $80$\% larger than that in ``$t^{H}$". 
This is because the self-energy $\Sigma ^{H}$ only weakly depends on the
wave number and orbital in contrast to $\Delta \Sigma ^{L}$. Therefore $\Sigma^H$ uniformly shrinks the band width from ``$t^{0}$". 

Since the downward shift of the $x^2-y^2$ orbital level in $t^0$ is even more enhanced than $t^H$,
we find that the essence of the shift is contained in the procedure of subtracting the exchange correlation potential of the LDA:
When the effect of the exchange correlation would be removed, the $x^2-y^2$ orbital would be located at a level much lower than other orbitals, because its stronger hybridization with the chalcogen $p$ orbitals contributes to the exchange correlation potentials.
By contrast, in FeTe, because the $yz/zx$ orbitals have the strong hybridization with the chalcogen $p$ orbitals (see Table \ref{spread_iron}), the reduction of the on-site potential of the $yz/zx$ orbitals is the largest.
Such weakly localized orbital, whose effective on-site interaction is relatively weak, has weaker self-energy compared to the other $3d$ orbitals.

More strictly speaking, the contribution of the $d$ electrons to the exchange correlation potential is more or less proportional to $U$ in amplitude and the sign is negative.
Therefore, if one removes this contribution, the level is pushed up more or less proportional to $U$.  Since $x^2-y^2$ orbital has the weakest $U$ for FeSe, the upward level shift is the smallest for $x^2-y^2$, meaning that it moves downward relative to other orbitals.
This interpretation is in accord with the interpretation above by the largest hybridization of the $x^2-y^2$ orbital, because the largest hybridization makes the largest Wannier spread and hence the weakest $U$.
The level shift is indeed in the order of $U$ both for FeSe and FeTe and in the order of the Wannier spread as well.
Then the level shift of $xy$, $yz/zx$, $3z^2-r^2$ and $x^2-y^2$ orbitals of FeSe from the LDA result to the result by ${\mathcal H}^{H\Delta}$ are $\sim +0.3, -0.05, +0.4$ and $-0.2$ eV, respectively, as we see in Table \ref{t_FeSe}.
Therefore, when we solve by using the present effective low-energy model with the single-particle part ${\mathcal H}^{H\Delta}$ instead of ${\mathcal H}^{\text{LDA}}$, the orbital levels should be effectively shifted with these amounts.

Since the correlation effect is governed by the most correlated $x^2-y^2$ orbital while the effective screened Coulomb interactions are nearly the same between the previous~\cite{miyake2} and the present models, such a downward shift of the $x^2-y^2$ orbital may increase the filling of the  $x^2-y^2$ orbital, and cause the depinning from half filling.
This may destroy the antiferromagnetic order as in the experimental observation.

Here, we note again that the band structure and the density of states shown in Figs. \ref{bndsFeSe}, \ref{DOSFeSe}, \ref{bndsFeTe} and \ref{DOSFeTe}
should not be taken as the properties that can be directly compared with the experiments,
because the experimentally accessible quantities such as the Fermi surface structure and the spectral weight are obtained only after solving the effective low-energy models.


\section{Summary}
We have proposed an improved scheme for \textit{ab initio} derivation of the low-energy effective models.
Our formalism is free from the double counting of Hartree and Fock contributions from the low-energy space, and the LDA exchange correlation is replaced by the GW self-energy.
Moreover, the derived effective model is reduced to a static one, where the $\omega$-dependence of the screened Coulomb interaction is taken into account and renormalized to the one-body part as a self-energy.
We have applied this formalism to transition-metal oxide SVO, as well as to iron-based superconductors FeSe and FeTe.
We have found there are two opposite effects, namely the increase in the band width arising from $\Sigma^{H}$ and the reduction arising from $\Delta \Sigma ^{L}$.
In SVO, these effects are roughly compensated, and remarkably,
the resultant band width and the dispersions are nearly the same as that of the LDA (the band width of LDA: $2.58$ eV, $\mathcal{H}^{H\Delta}$: $2.56$ eV).
On the other hand, in the non-degenerate multi-band systems such as FeSe and FeTe, though the band widths are also similar to the LDA results, the momentum and orbital dependent self-energy effects yield modifications of the resultant band structures.
For the effective low energy model for FeSe given by $\mathcal{H}^{H\Delta}$, the on-site potential of the $x^{2}-y^{2}$ orbitals, which has the strongest hybridization with the Se $4p$ orbitals and thus the weakest on-site interaction among the Fe $3d$ orbitals, is substantially lowered from that of the LDA.
This may make the occupation number of the $x^{2}-y^{2}$ orbital away from half filling and destroy the antiferromagnetic order as in the experimental observation.
In contrast, the lowering of the on-site potentials is the largest in the $yz/zx$ orbitals in FeTe, and the resultant occupation number of the $x^{2}-y^{2}$ orbital remains close to half filling.

By our formalism, two major drawbacks in the derivation of the low-energy effective model in the literature are removed. 
The effects of $\Sigma^{H}$ and $\Delta \Sigma ^{L}$ compensate and the resultant band width is, rather accidentally, nearly the same as that of the LDA, while the structure of the band and the on-site potential of the nonequivalent orbital quantitatively change in some systems such as FeSe and FeTe.

Even with this improvement, we still have future issues to be resolved: In the practical calculation of the low-energy effective model by low-energy solvers, one often neglects the farther-neighbor effective interaction owing to heavy costs of the numerical computation. This long-range part makes the effect of the on-site interaction weaker.
The polaronic screening effect from the lattice also weakens the effective Coulomb interaction between the electrons. When simplified models are used in the low-energy solvers, these correction may have small contributions.
Counting of these effects to the downfolding formalism are left for future problems.

Another open issue is to consider possible refined estimates of the GW self-energy arising from the high-energy contributions.
A self-consistent GW scheme instead of the one-shot self-energy has been employed by Kutepov {\it et al.}~\cite{Kutepov} with the combination of the solver based on the  dynamical mean field theory.
In the full GW calculations, it is known for a long time that the self-consistent GW calculations give worse agreements with the experimental estimates of the gap amplitudes than those of the one-shot GW.
This may be attributed to the vertex corrections ignored in the GW approximation that roughly cancels the difference between the one-shot and self-consistent estimates.
Counting both of the vertex corrections and self-consistency is, though expected to cause minor differences in this constrained self-energy, left for future studies.


\section{Acknowledgments}
We would like to thank Kazuma Nakamura for discussions on the downfolding method and the double counting of the correlation
in the low-energy effective model, and Silke Biermann for discussions on $\omega$-dependence of the screened Coulomb interaction.
MH would like to thank Takahiro Misawa for useful advice and fruitful discussions on iron-based superconductors.
This work has been supported by Grants-in-Aid for Scientific Research from the Ministry of Education, Culture, Sports,
Science and Technology of Japan (MEXT) under the grant numbers 22104010 and 22340090.
This work has also been financially supported by MEXT HPCI Strategic Programs for Innovative Research (SPIRE) and Computational
Materials Science Initiative (CMSI).


\end{document}